\documentclass[a4,11pt,draftcls,onecolumn]{IEEEtran}

\usepackage{graphicx}
\usepackage{epstopdf}
\usepackage{float}
\usepackage{subfig}

\usepackage{cite}
\usepackage{amsmath, amssymb, mathbbol}
 
\renewcommand{\QED}{\QEDopen}

\usepackage[active]{srcltx} 
\usepackage{color}
\usepackage{rotating}
\usepackage[colorlinks=true,linkcolor=blue,citecolor=blue]{hyperref}

\begin{document}
\title{On Universal D-Semifaithful Coding for\\[-2mm] Memoryless Sources with Infinite Alphabets }
\author{Jorge F. Silva,~\IEEEmembership{Senior Member,~IEEE} and Pablo Piantanida,~\IEEEmembership{Senior Member,~IEEE}
\thanks{The material in this paper was partially published in the Proceedings  of the 2019 IEEE International Symposium on Information Theory (ISIT)~\cite{silva_2019_isit}.}
\thanks{J. F. Silva is with the Information and Decision Systems (IDS) Group,  University of Chile, Av. Tupper 2007 Santiago, 412-3, Room 508, Chile, Tel: 56-2-9784090, Fax: 56-2 -6953881,   (email: josilva@ing.uchile.cl).}
\thanks{P. Piantanida is with the Laboratoire des Signaux et Syst\`emes (L2S), CentraleSup\'elec, CNRS, Universit\'e Paris-Saclay, France and with the Montreal Institute for Learning Algorithms (Mila), Canada  (email: pablo.piantanida@centralesupelec.fr).}
}
\maketitle

\begin{abstract}
The problem of variable length and fixed-distortion universal source coding (or D-semifaithful source coding) for stationary and memoryless sources on countably infinite alphabets ($\infty$-alphabets) is addressed in this paper.  The main results of this work offer a set of sufficient conditions (from weaker to stronger) to obtain weak minimax universality,  strong minimax universality, and corresponding achievable rates of convergences for the worse-case redundancy for the family of stationary memoryless sources whose densities are dominated by an envelope function (or the envelope family) on $\infty$-alphabets. An important implication of these results is that universal D-semifaithful source coding is not feasible for the complete family of stationary and memoryless sources on $\infty$-alphabets.  To demonstrate this infeasibility,  a sufficient condition for the impossibility is presented for the envelope family. Interestingly, it matches the well-known impossibility condition in the context of  lossless (variable-length) universal source coding. More generally, this work offers a simple description of what is needed to achieve universal  D-semifaithful coding for a family of distributions $\Lambda$. This reduces to finding a collection of quantizations of the product space at different block-lengths --- reflecting the fixed distortion restriction --- that satisfy two asymptotic requirements: the first is a universal quantization condition with respect to $\Lambda$,  and the second is a vanishing information radius (I-radius) condition for $\Lambda$ reminiscent of the condition known for  lossless universal source coding. 
\end{abstract}

\begin{keywords}
Lossy  compression, variable length source coding, $D$-semifaithful code, universal source coding, infinite alphabets, strong minimax universality,  information radius, universal quantization, envelope families. 
\end{keywords}

 \newtheorem{corollary}{\bf COROLLARY}
\newtheorem{theorem}{\bf THEOREM}
\newtheorem{lemma}{\bf LEMMA}
\newtheorem{proposition}{\bf PROPOSITION}
\newtheorem{definition}{\bf Definition}
\newtheorem{remark}{\bf Remark}

\section{Introduction}
\label{sec_intro}
Universal Source Coding (USC) has a long history \cite{csiszar_2004,cover_2006,gyorfi_1994,davisson_1973,kieffer_1978},  
starting with the seminal work of Davisson~\cite{davisson_1973} who formalized the variable-length lossless coding problem and introduced relevant information quantities.  In lossless variable-length source coding,  it is well-known  that if we know the statistics of a stationary and memoryless source, the Shannon entropy of the 1D marginal of the process characterizes the minimum achievable rate~\cite{cover_2006}. However, when the statistics of the source are not known  but the source belongs to a family of stationary and memoryless distributions $\Lambda$, the problem reduces to characterizing the worst-case expected overhead  (or \emph{worse-case redundancy}) that a pair of encoder and decoder exhibit due to the lack of knowledge about true distribution~\cite{csiszar_2004,gassiat_2018}. In fact, a seminal information-theoretic result states that the least worst-case overhead (or minimax redundancy of $\Lambda$) is fully characterized by the \emph{information radius} of $\Lambda$~\cite{csiszar_2004}.  

The information radius (I-radius) has been richly studied by the community, and there are numerous contributions~\cite{boucheron_2009,bontemps_2014,haussler_1997,54897, 382017}. In particular, it is well-known that the I-radius grows sub-linearly for the family of finite alphabet stationary and memoryless sources~\cite{csiszar_2004}, which implies the existence of a {universal source code} that achieves {Shannon entropy} for every distribution in this family provided that the block length tends to infinity.  Unfortunately,  this positive result does not extend to the case of stationary and memoryless sources on {\em countably infinite alphabets} ($\infty$-alphabets)~\cite{kieffer_1978,gyorfi_1994,boucheron_2009}. From an information complexity perspective,  this infeasibility result means that the I-radius of this family is unbounded for any finite block-length; consequently,  lossless universal source coding for $\infty$-alphabet stationary and memoryless sources is an intractable problem.

There has been renewed interest in USC with infinite alphabets in recent year \cite{boucheron_2009,bontemps_2014, boucheron_2015,silva_2017,silva_piantanida_reprint_2017}.
Restricting the study to the case of memoryless sources with marginal densities dominated by an envelope function $f$ (or the envelope family $\Lambda_f$),  a series of new  results have been presented in~\cite{boucheron_2009,bontemps_2014,boucheron_2015, silva_piantanida_reprint_2017}. Remarkably,  \cite[Theorems 3 and 4]{boucheron_2009} show that $f$ being summable (over the infinite alphabet) is a necessary and sufficient condition to guarantee strong minimax universality for the envelope family $\Lambda_f$.  Consequently, universality can be achieved for a non-trivial (infinite dimensional) collection of distributions with infinite support. Furthermore,  the specific rate of convergence for the worse-case redundancy (i.e., the information radius of $\Lambda_f$) has been derived for exponential and power law (envelope) families in $\infty$-alphabets as well the construction of coding schemes that achieve optimal worse-case redundancies (information limits) \cite{bontemps_2014,boucheron_2015}, among other interesting results.  

Complementing the previous results on infinity alphabet sources and using ideas from weak source coding by Han~\cite{han_2000},  the almost lossless universal source coding was introduced in \cite{silva_piantanida_reprint_2017,silva_isit_2016}. The general of this approach is to relax the lossless assumption by introducing a non-zero distortion that tends to zero with the block-length (asymptotic zero distortion), with the intention of achieving weak universality over the entire collection of memoryless sources on $\infty$-alphabets \cite{silva_2017,silva_piantanida_reprint_2017}. Results in this weak setting demonstrate that almost lossless USC is feasible for the entire family of stationary and memoryless distributions \cite[Th. 4]{silva_piantanida_reprint_2017} on $\infty$-alphabets, and the sensitive role that 
 the vanishing distortion plays on the analysis of the problem when moving from a point-wise to a uniform convergence to zero \cite[Th. 5]{silva_piantanida_reprint_2017}. 

\subsection{Contributions}
In this paper,  we investigate the problem introduced by Ornstein and Shields in \cite{ornstein_1990} of fixed-distortion and variable length universal source coding---or universal $D$-semifaithful coding---for $\infty$-alphabet sources. Following  the line of work of the seminal paper by Boucheron \emph{et al.}~\cite{boucheron_2009}, among others~\cite{bontemps_2014,boucheron_2015,silva_piantanida_reprint_2017},  we study the family of stationary and memoryless sources whose densities are dominated by an envelope function $f$ by adopting the criterion of strong minimax universality~\cite{csiszar_2004}. The redundancy in this case is measured with respect to the rate-distortion function lower bound \cite{berger_1971,cover_2006,gray_1990}. Our main results (cf. Theorem~\ref{main_th_part1} and \ref{main_th_rate_overhead_envelope}) parallel the results presented in the lossless problem \cite[Theorems 3 and 4]{boucheron_2009}  and offer a set of conditions on the envelope function to obtain weak minimax universality,  strong minimax universality  as well as  an achievable rate of convergence for the worse-case redundancy. Conversely, Theorem \ref{main_th_part1} shows that if the envelope function is not summable, then strong minimax universality is not feasible, i.e., an impossible result. Indeed, this result matches the infeasibility condition known for the case of lossless USC~\cite{boucheron_2009}. More generally, we present a simple result  that captures what is needed (necessary and sufficient conditions) to achieve universal $D$-semifaithful  source coding in terms of some asymptotic properties imposed on a collection of partitions of the source alphabet (Theorem \ref{th_necessary_suffuficient_cond_universality}).

A central technical contribution of this paper relies on the derivation of a lower bound for the minimax redundancy of a $D$-semifaithful code, operating at a given distortion level, which is obtained using a redefined expression of the I-radius for the family of sources. The resulting I-radius expression is based on the information divergence restricted to quantization cells (or bins) induced by the $D$-semifaithful code. This lower bound represents the central ingredient to derive the impossibility argument over envelope families. On the other hand, achievable results are obtained for summable envelope functions, similarly to the case of lossless source coding~\cite{boucheron_2009,bontemps_2014}. For this a two-stage constructive coding scheme is employed (operating at a fixed distortion)  for which results are adopted from universal $D$-semifaithful coding on finite alphabets (Lemma \ref{lemma_finite_alphabet_unievrsal_construction}) and universal lossless source coding on $\infty$-alphabets \cite{boucheron_2009,bontemps_2014}. To the best of our knowledge, our results are the first that explore universal $D$-semifaithful coding for stationary and memoryless sources on $\infty$-alphabets using the criterion of strong minimax universality.  A preliminary version of this paper was presented in \cite{silva_2019_isit} where some of the results were introduced without a complete presentation of their proofs.

\subsection{Related Work on Universal $D$-semifaithful for Finite Alphabet Sources}
Relevant results on universal $D$-semifaithful coding have been presented for finite alphabet sources~\cite{ornstein_1990,yu_1993,kontoyiannis_2000}.  In particular,  Ornstein and Shields \cite{ornstein_1990} proposed a universal $D$-semifaithful code for finite alphabet ergodic sources deriving almost-sure convergence of the rate of the code to the rate-distortion function (a sample-wise analysis). 
Complementing this analysis, Yu and Speed \cite{yu_1993} proposed a two-stage universal D-semifaithful code 
for the family of finite alphabet stationary and memoryless sources with some added regularity conditions.  They showed that the average rate of this  $D$-semifaithful code achieves (uniformly over this family) the rate-distortion function at a rate of convergence that is $O(n^{-1}\log n)$. On the optimality of this last constructive result,  it is showed in \cite{merhav_1995}  that the rate $O(n^{-1}\log n)$ is optimal at least for the {\em Hamming} distortion measure.  This optimality  was showed more generally in \cite{zhang_1997} and they also presented new schemes that achieve the optimal rate of convergence of $O(n^{-1}(\log n+ o(\log n)))$ for finite alphabet stationary and memoryless sources.  Results of the same nature were obtained in \cite{ishii_1997}.  Revisiting the sample-wise redundancy analysis  of lossy source coding operating at a fixed distortion,  Kontoyiannis \cite{kontoyiannis_2000} showed that the best (sample-wise) redundancy rate (in bits per sample) of a code that knows the model is $O(1/\sqrt{n})$ (a converse result).  The analysis was then extended to a universal setting, where for finite alphabet memoryless sources the same redundancy rate (sample-wise) of  $O(1/\sqrt{n})$ is shown. Surprisingly in terms of sample wise redundancy, this work showed that no penalization is observed when moving from an optimal code that knows the model to a universal setting for finite alphabet memoryless sources.  This matching is non-observed when the analysis is based on the average redundancy of a code \cite{csiszar_2004}.

The rest of the paper is organized as follows.  Section \ref{sec_pre} introduces some  definitions and basic elements for the formalization of the problem. Section \ref{sub_sec_UdSC} presents the universal D-Semifaithful source coding problem and introduces a general result (Theorem \ref{th_necessary_suffuficient_cond_universality}).  Section \ref{sec_main} presents results for the family of envelope distributions (Theorems \ref{main_th_part1} and \ref{main_th_rate_overhead_envelope}).  The  arguments used to prove the main results, Theorems \ref{main_th_part1} and \ref{main_th_rate_overhead_envelope} are presented in Section \ref{sec_proof_main}. 
Final remarks and directions for future work are presented in Section \ref{final}. Finally, supporting results and technical derivations are relegated to the Appendix sections.

\section{Main Definitions and Preliminaries}
\label{sec_pre}
Let us denote by $\mathbb{X}$ a countably infinite alphabet, without loss of generality the integers.  
The space is equipped with a distortion function $\rho:\mathbb{X} \times \mathbb{X} \longrightarrow \mathbb{R}^+$, and  
the non-trivial scenario is assumed where $\rho(x,\bar{x})>0$ if $\bar{x}\neq x$.  For any $n\geq 1$, we have 
$\rho_n: \mathbb{X}^n \times \mathbb{X}^n \longrightarrow \mathbb{R}^+$  of block length $n$ to be the standard single letter construction obtained from $\rho$  \cite{berger_1971,gray_1990}, where for any $x^n=(x_1,..,x_n)$ and  $\bar{x}^n=(\bar{x}_1,..,\bar{x}_n)$  in $\mathbb{X}^n$
\begin{equation}\label{eq_sec_pre_0}
\rho_n(x^n, \bar{x}^n) \equiv \frac{1}{n}\sum_{i=1}^n \rho(x_i,\bar{x}_i).
\end{equation}

A $D$-semifaithful code of length $n$ operating at a distortion $d>0$ is a variable length coding scheme operating 
at a fixed distortion \cite{ornstein_1990,kontoyiannis_2000}.  More precisely,  we consider the following definition:
\begin{definition}
\label{def_semifaithful_code}
A  $D$-semifaithful code of length $n$ operating at distortion $d>0$ is defined/denoted by a triplet $\xi_n=(\phi_n, \mathcal{C}_n, \mathcal{D}_n)$,  where 
\begin{itemize} 
	\item $\phi_n: \mathbb{X}^n \longrightarrow \mathcal{B}_n \subset \mathbb{X}^n$ is a quantizer, 
	\item $\mathcal{C}_n:  \mathcal{B}_n \longrightarrow \left\{0,1 \right\}^{*} \equiv  \cup_{k\geq 1} \left\{0,1 \right\}^k$ is a binary (variable length and prefix-free) encoder, and
	\item $\mathcal{D}_n:  \left\{0,1 \right\}^{*}   \longrightarrow \mathcal{B}_n$ is a binary decoder, 
\end{itemize}
satisfying that for any $x^n\in \mathbb{X}^n$ 
\begin{equation}\label{eq_sec_pre_0b}
\rho_n(x^n, \phi_n(x^n))\leq d. 
\end{equation}
The set $\mathcal{B}_n=\left\{ \phi_n(x^n), x^n \in \mathbb{X}^n \right\}$ contains the prototypes of $\xi_n$ in $\mathbb{X}^n$. In this construction, the binary encoder $\mathcal{C}_n$, which is variable length, is prefix-free \cite{cover_2006} meaning that it satisfies the {\em  Kraft-MacMillan inequality}: 
$$\sum_{i\in  \mathcal{B}_n} 2^{-\mathcal{L}(\mathcal{C}_n(i))} \leq 1,$$ 
where $\mathcal{L}:\left\{0,1 \right\}^{*} \longrightarrow \mathbb{N}\setminus \left\{ 0\right\}$ is the function that returns the length (number of bits) of a vector in $\left\{0,1 \right\}^{*}$. 
\end{definition}

Importantly for the analysis presented in this paper,  the code $\xi_n=(\phi_n, \mathcal{C}_n, \mathcal{D}_n)$ induces a partition in $\mathbb{X}^n$ given/denoted by 
\begin{equation}\label{eq_sec_pre_0c}
\pi_{\phi_n}\equiv \left\{ \mathcal{A}_{n,y^n} \equiv \phi^{-1}_n(\left\{y^n \right\}), y^n \in  \mathcal{B}_n \right\}\subset 2^{\mathbb{X}^n}, 
 \end{equation}
where we assume the non-suboptimal (and expected) condition that $y^n \in \mathcal{A}_{n,y^n}$ for any prototype $y^n\in \mathcal{B}_n$. 

\subsection{The Source Coding Problem}
\label{sub_sec:source_coding_basic}
Let us consider an information source (a random sequence) $X=(X_n)_{n\geq 1}$ with values in $\mathbb{X}$ and
process distribution denoted by $\mu=\left\{\mu_n\in \mathcal{P}(\mathbb{X}^n), n\geq 1 \right\}$, where for any $n\geq 1$ $X^n=(X_1,..,X_n)\sim \mu_n$, and $\mathcal{P}(\mathbb{X}^n)$ denotes the collection of probabilities in $\mathbb{X}^n$. Then, the rate (in bits per sample) for encoding $X^n$ with a $D$-semifaithful code $\xi_n$ of length $n$ operating at distortion $d>0$ is given  by
\begin{equation}\label{eq_sec_pre_1}
	R(\xi, \mu_n) \equiv \frac{1}{n} \mathbb{E}_{X^n\sim \mu_n}  \left\{  \mathcal{L}(\mathcal{C}_n(\phi_n(X^n))) \right\}.
\end{equation}

Using the source model $\mu$, 
the variable length fixed distortion lossy source coding problem reduces to minimizing  $R(\xi, \mu_n)$  in  (\ref{eq_sec_pre_1}) 
over the family of $D$-semifaithful codes (operating at distortion $d$) for any $n\geq 1$ \cite{kieffer_1978,kieffer_1991}. 
It is well-known  that for any $D$-semifaithful code $\xi_n$ \cite{cover_2006,csiszar_2004}
\begin{equation}\label{eq_sec_pre_1b}
n R(\xi, \mu_n) \geq H(v_{\mu_n}),  
\end{equation}
where $v_{m_n}$ denotes the probability induced by $\mu_n$ and $\phi_n$ in the reproducible alphabet $\mathcal{B}_n$, i.e.,  $v_{m_n}(y^n)=\mu_n(\phi^{-1}_n(\left\{ y^n \right\}))$ for any $y^n\in \mathcal{B}_n$, and 
\begin{equation}\label{eq_sec_pre_1c}
H(v_{\mu_n}) \equiv - \sum_{i\in  \mathcal{B}_n } v_{\mu_n}(i) \log (v_{\mu_n}(i))
\end{equation} 
is the {\em Shannon entropy} of $v_{\mu_n} \in \mathcal{P}(\mathcal{B}_n)$ \cite{cover_2006,gray_1990} and the $\log$ function is base $2$.  Furthermore, fixing $\phi_n$ (the quantizer) and optimizing over the encoder-decoder pairs $(\mathcal{C}_n, \mathcal{D}_n)$ (the prefix-free mappings from $\mathcal{B}_n$ to $\left\{0,1 \right\}^{*}$), we have that \cite{cover_2006,gray_1990}:
\begin{equation}\label{eq_sec_pre_2}
	\frac{H(v_{\mu_n})+ 1}{n} \geq \min_{(\mathcal{C}_n, \mathcal{D}_n)} R( (\phi_n,\mathcal{C}_n, \mathcal{D}_n), \mu_n) \geq \frac{H(v_{\mu_n})}{n}.
\end{equation}

A  convenient way to write the entropy of the induced distribution $v_{m_n}$ in (\ref{eq_sec_pre_2}) is as the entropy of $\mu_n$ but projected over quantization (or a sub-sigma field of the measurable space $(\mathbb{X}^n, 2^{\mathbb{X}^n})$).  Given a partition $\pi=\left\{A_i, i\in \mathcal{I} \right\}$ (countable or finite) of $\mathbb{X}^n$ and a probability $\mu\in \mathcal{P}(\mathbb{X}^n)$, we introduce the entropy of $\mu$ restricted over the sub-sigma field $\sigma(\pi)$ 
by:
\begin{align}\label{eq_sec_pre_3}
	H_{\sigma(\pi)}(\mu) &\equiv - \sum_{i \in \mathcal{I}}  \mu(A_i) \log \mu(A_i)  \leq H(\mu)\nonumber\\
	&=- \sum_{x^n\in \mathbb{X}^n} \mu(x^n)\log \mu(x^n),
\end{align}
where the last inequality follows from basic information inequalities \cite{cover_2006}.  Then,  $H(v_{\mu_n})$
is equal to $H_{\sigma(\pi_{\phi_n})}(\mu_n)$ 
and (\ref{eq_sec_pre_2}) can be re-written by: 
\begin{equation}\label{eq_sec_pre_4}
	\frac{H_{\sigma(\pi_{\phi_n})}(\mu_n)+ 1}{n} \geq \min_{(\mathcal{C}_n, \mathcal{D}_n)} R( (\phi_n,\mathcal{C}_n, \mathcal{D}_n), \mu_n) \geq \frac{H_{\sigma(\pi_{\phi_n})}(\mu_n)}{n}.
\end{equation}

From (\ref{eq_sec_pre_4}), the source coding (operational) problem is 
\begin{equation}\label{eq_sec_pre_5a}
 R_n(d,\mu_n) \equiv \min_{\xi_n} R(\xi, \mu_n), 
 \end{equation}
where $\xi_n$ is running over the family of $D$-semifaithful codes of length $n$ operating at distortion $d$ (Def.\ref{def_semifaithful_code}). 
This operational problem can be considered 
equivalent to solve\footnote{Up to a discrepancy of at most $1/n$ in bits per sample.}:
\begin{equation}\label{eq_sec_pre_5b}
	\mathcal{R}_n(d,\mu_n) \equiv \min_{\pi \in \mathcal{Q}_n(d)}  \frac{H_{\sigma(\pi)}(\mu_n)}{n},
\end{equation}
where $\mathcal{Q}_n(d)$ denotes the collection of partitions of $\mathbb{X}^n$ where any $\pi$  in  $\mathcal{Q}_n(d)$
satisfies that:  $\forall \mathcal{A}\in \pi$, $\exists y^n \in \mathcal{A}$ such that 
$$\sup_{x^n\in A}\rho_n(x^n,y^n) \leq d,$$
 i.e., any $\pi \in \mathcal{Q}_n(d)$ offers a $d$-covering of $\mathbb{X^n}$ with respect to $\rho_n$.
 
For memoryless and stationary sources, it is well known that 
$\lim_{n \rightarrow \infty }\mathcal{R}_n(d,\mu_n)$  convergences 
to the celebrated rate-distortion function \cite{cover_2006,gray_1990}, which is a function of $\mu_1\in \mathcal{P}(\mathbb{X})$ \cite{kieffer_1978,kieffer_1991}.  For completeness, we briefly revisit this result here.

\subsection{The Source Coding Theorem}
\label{sub_sec:source_coding_th}
Let us consider  $(X_n)_{n\geq 1}$ to be a stationary and memoryless source characterized by $\mu_1\in \mathcal{P}(\mathbb{X})$.  
The rate distortion function of $\mu= \left\{\mu_n, n\geq 1 \right\}$ relative to $\rho$ is given by \cite{kieffer_1978}:  
	$$\inf_{n \geq 1} \mathcal{R}^*(d,\mu_n) = \lim_{n \rightarrow \infty} \mathcal{R}^*(d,\mu_n),$$
where
\begin{equation}\label{eq_sec_pre_7}
	\mathcal{R}^*(d,\mu_n) \equiv \frac{1}{n} \inf_{\mathbf{U},\mathbf{V}}  I(\mathbf{U};\mathbf{V}).
\end{equation}
The infimum in (\ref{eq_sec_pre_7}) is taken with respect to the collection of 
joint random vectors $(\mathbf{U},\mathbf{V})$ in $\mathbb{X}^n\times \mathbb{X}^n$ satisfying that 
$U\sim \mu_n$ and 
$\mathbb{P}(\rho_n(\mathbf{U},\mathbf{V}) \leq d)=1$ \cite{kieffer_1978}.
By the definitions of these objects, it is simple to verify that $R_n(d,\mu_n) \geq \mathcal{R}_n(d,\mu_n) \geq \mathcal{R}^*(d,\mu_n)$
for any $n\geq 1$.  Importantly,  Kieffer showed that:

\begin{theorem}(Kieffer \cite[Th. 4]{kieffer_1978})
\label{main_source_coding_theorem}
 For a D-semifaithful coding problem operating at distortion $d>0$,  
\begin{equation}\label{eq_sec_pre_7b}
\lim_{n \rightarrow \infty }\mathcal{R}_n(d,\mu_n)
=\lim_{n \rightarrow \infty} \mathcal{R}^*(d,\mu_n)= \mathcal{R}^*(d,\mu_1).
\end{equation}
\end{theorem}
The last expression in (\ref{eq_sec_pre_7b}) is the single letter information theoretic limit of this problem \cite{kieffer_1978}.

\section{A General Result on Universal D-Semifaithful Coding}
\label{sub_sec_UdSC}
In universal source coding, the objective is to find a coding scheme that achieves the 
performance limit in (\ref{eq_sec_pre_7b})
without  knowledge of the underlying source distribution \cite{gassiat_2018,csiszar_2004}.
To formalize this problem in the context of D-semifaithful coding, 
let $(X_n)_{n\geq 1}$ be a stationary and memoryless source with values in $\mathbb{X}$,
where we impose that $\mu_1$ belongs to $\Lambda \subset \mathcal{P}(\mathbb{X})$. 
Let  $\left\{\xi_n=(\phi_n, \mathcal{C}_n, \mathcal{D}_n), n\geq 1 \right\}$  be a D-semifaithful coding scheme operating 
at distortion $d>0$ with respect to the single letter distortions $\left\{ \rho_n, n\geq 1\right\}$. 
Following the definitions used in universal lossless source coding  \cite{davisson_1973}, we say that: 
\begin{definition}
\label{def_strong_minimax_universality}
A coding scheme $\left\{\xi_n,  n\geq 1 \right\}$  (operating  at distortion $d>0$) is strongly minimax universal for $\Lambda$
at distortion $d$ if,
\begin{equation}\label{eq_sub_sec_UdSC_1}
	\lim_{n \rightarrow \infty} \underbrace{\sup_{\mu^n \in \Lambda^n}   \left[ R(\xi_n, \mu^n) - \mathcal{R}_n(d,\mu^n) \right]  }_{\text{worse-case redundancy over } \Lambda^n \text{ of } \xi_n}=0, 
\end{equation}
where $\Lambda^n \equiv \left\{\mu^n, \mu\in \Lambda \right\} \subset \mathcal{P}(\mathbb{X}^n)$, and $\mu^n$ is the product (i.i.d.)  distribution induced by $\mu\in \mathcal{P}(\mathbb{X})$.  
\end{definition}

By definition of $\mathcal{R}_n(d,\mu^n)$ in (\ref{eq_sec_pre_5b}), we have that $R(\xi_n, \mu^n) - \mathcal{R}_n(d,\mu^n) \geq 0$ and, consequently,  this last expression can be interpreted as the redundancy (in bits per sample) we have to accept for not knowing the distribution of $X^n$ and using 
a distribution independent lossy encoder.  Therefore if $\left\{\xi_n,  n\geq 1 \right\}$ is strongly minimax universal, it means that as the block length tends to infinity  (and uniformly over the family of hypotheses in $\Lambda$),  the scheme achieves the best performance obtained by a scheme that knows the distribution of the source  previous to encoding.   
Similarly,   we say that:
\begin{definition}
\label{def_weak_minimax_universality}
A scheme $\left\{\xi_n,  n\geq 1 \right\}$ (operating  at distortion $d>0$) is weakly minimax universal for $\Lambda$
at distortion $d$  if \cite{davisson_1973},
\begin{equation}\label{eq_sub_sec_UdSC_2}
	\lim_{n \rightarrow \infty}   \left[ R(\xi_n, \mu^n) - \mathcal{R}_n(d,\mu^n) \right] =0, \ \ \forall \mu\in \Lambda. 
\end{equation}
\end{definition}
In contrast to Definition \ref{def_strong_minimax_universality}, being weakly minimax universal imposes a point-wise convergence of the redundancy over the collection of hypotheses in $\Lambda$.

Before we move to the presentation of the main context of study of this work, we present a general analysis for the worse-case redundancy in (\ref{eq_sub_sec_UdSC_1}).  
\subsection{Minimax Redundancy Analysis}
\label{sec_proof_main_pre}
Let $\xi_n=(\phi_n, \mathcal{C}_n, \mathcal{D}_n)$ be a $D$-semifaithful code of length $n$ operating at distortion $d>0$,  and $\mu$ be a distribution in $\Lambda \subset \mathcal{P}(\mathbb{X})$.  Then, the average redundancy of $\xi_n$ (in bits per sample) can be expressed by
\begin{equation}\label{eq_main_proof_pre_1}
	R(\xi_n, \mu^n) - \mathcal{R}_n(d,\mu^n) = \left[ R(\xi_n, \mu^n)  - \frac{H_{\sigma(\pi_{\phi_n})}(\mu^n)}{n}   \right]  + \left[ \frac{H_{\sigma(\pi_{\phi_n})}(\mu^n)}{n}  - \mathcal{R}_n(d,\mu^n) \right],
\end{equation}
where $\pi_{\phi_n}$ is the partition of $\mathbb{X}^n$ induced by $\phi_n$ (see Eq.(\ref{eq_sec_pre_0c})), and $\mu^n=\mu \times ..\times \mu\in \mathcal{P}(\mathbb{X}^n) $ is a short-hand 
for the $n$-fold distribution induced by $\mu$.  In particular, the first term on the right-hand-side (RHD) of (\ref{eq_main_proof_pre_1}) is non-negative from (\ref{eq_sec_pre_4}) and the second term is non-negative from the definition in (\ref{eq_sec_pre_5b}). 
\subsubsection{The Projected Information Radius of $\Lambda^n$ with Respect to $\pi_{\pi_n}$} \label{sec_proof_main_pre_information_radius}
For the moment, let us concentrate on the analysis of $\left[ R(\xi, \mu^n)  - {H_{\sigma(\pi_{\phi_n})}(\mu^n)}/{n}   \right]$ in (\ref{eq_main_proof_pre_1}). From a well-known connection between distributions and prefix-free codes \cite{cover_2006}, 
the encoder $\mathcal{C}_n$ can be associated with a distribution $v_{\mathcal{C}_n} \in \mathcal{P}(\mathcal{B}_n)$
and $R(\xi, \mu^n) - {H_{\sigma(\pi_{\phi_n})}(\mu^n)}/{n} $ can be approximated (up to a discrepancy of $1/n$) by 
$$\frac{1}{n} D(v_{\mu^n}\|v_{\mathcal{C}_n})= \frac{1}{n}  \sum_{y^n \in \mathcal{B}_n} v_{\mu^n}(y^n) \log \frac{v_{\mu^n}(y^n)}{v_{\mathcal{C}_n}(y^n)} \geq 0,$$ 
where $v_{\mu^n} \in \mathcal{P}(\mathcal{B}_n)$ is a short-hand for
the distribution induced by $\mu^n$ and $\phi_n$ in the reproducible space $\mathcal{B}_n$.  Then,  we can consider 
the worse case (over $\Lambda$) of this discrepancy by
\begin{equation}\label{eq_main_proof_pre_2}
	R^+_n(\Lambda,\underbrace{\xi_n}_{(\phi_n, \mathcal{C}_n, \mathcal{D}_n)}) \equiv \frac{1}{n}  \sup_{\mu \in \Lambda} D(v_{\mu^n}\|v_{\mathcal{C}_n})\geq 0. 
\end{equation}
For the rest of the analysis, it is convenient to fix the quantization $\phi_n$ (i.e., $\mathcal{B}_n$ and its associated partition $\pi_{\phi_n}$) and optimize the prefix-free mapping from $\mathcal{B}_n$ to $\left\{0,1 \right\}^*$ with respect to the 
divergence term in  (\ref{eq_main_proof_pre_2}). The solution of this problem introduces the information radius of the family $\Lambda^n$ projected over the sigma field induced by the partition  $\pi_{\phi_n}$ \cite{csiszar_2004}.  More precisely, we obtain the following:
\begin{align}\label{eq_main_proof_pre_3}
	\min_{(\mathcal{C}_n, \mathcal{D}_n)}  \sup_{\mu \in \Lambda} \left[ R(\xi_n=(\phi_n, \mathcal{C}_n, \mathcal{D}_n), \mu^n)  - \frac{H_{\sigma(\pi_{\phi_n})}(\mu^n)}{n}   \right]   &\approx  \min_{(\mathcal{C}_n, \mathcal{D}_n)} R^+_n(\Lambda, {\xi_n=(\phi_n, \mathcal{C}_n, \mathcal{D}_n)})\\
	\label{eq_main_proof_pre_3a}
	&= \frac{1}{n} R^+(\Lambda^n, \sigma(\pi_{\phi_n})),
\end{align}
where from (\ref{eq_main_proof_pre_2})
\begin{align}   \label{eq_main_proof_pre_3b}
R^+(\Lambda^n, \sigma(\pi_{\phi_n}))  &\equiv  \min_{v\in \mathcal{P}(\mathcal{B}_n)}  \sup_{\mu^n \in \Lambda^n} D(v_{\mu^n}\|v)\nonumber\\
	   &= \underbrace{ \min_{v\in \mathcal{P}(\mathbb{X}^n)}  \sup_{\mu^n \in \Lambda^n}  D_{\sigma(\pi_{\phi})} (\mu^n \|v)}_{\text{information radius of $\Lambda^n$ projected on $\pi_{\phi_n}$}}.
 \end{align}
The last expression in (\ref{eq_main_proof_pre_3b}) is written in terms of the divergence between distributions on the original sample space $\mathbb{X}^n$ but restricted  over the cells of $\pi_{\phi_n}$  using that:  
\begin{align}\label{eq_main_proof_pre_3c}
D_{\sigma(\pi)}(\mu\|v) \equiv \sum_{A\in \pi} \mu(A) \log \frac{ \mu(A) }{ v(A) }\leq D(\mu\|v), 
\end{align}
for any $\pi$ partition of $\mathbb{X}^n$ and $\mu,v\in \mathcal{P}(\mathbb{X}^n)$. 
Finally, the approximation in (\ref{eq_main_proof_pre_3}) is up to a discrepancy of  $1/n$. 

In summary for a fixed quantizer $\phi_n$, optimizing the second-stage
(over the collection of prefix-free encoder-decoder pairs) reduces to the information radius problem  in (\ref{eq_main_proof_pre_3b}). This problem finds the distribution that is closest to the entire family $\Lambda_f^n$ (or the centroid of the family) using the divergence
restricted over the sub-sigma field $\sigma(\pi_{\phi_n})$ in (\ref{eq_main_proof_pre_3c}).
Interestingly, this is the same information radius characterization used in universal (variable length) lossless source coding \cite{csiszar_2004}.

\subsubsection{Universal Quantization over $\Lambda^n$} Let us now concentrate on the analysis of the other term 
$$\left[ {H_{\sigma(\pi_{\phi_n})}(\mu^n)}/{n}  - \mathcal{R}_n(d,\mu^n) \right]$$ 
in (\ref{eq_main_proof_pre_1}), which depends exclusively on the quantizer $\phi_n$ (or 
equivalently on $\pi_{\phi_n} \in \mathcal{Q}_n(d)$,  see (\ref{eq_sec_pre_5b})). Then moving to the universal setting, 
it is reasonable to optimize $\pi_{\phi_n} \in \mathcal{Q}_n(d)$ over the worse-case discrepancy given by: 
\begin{align}\label{eq_main_proof_pre_4}
	 \min_{\bar{\pi} \in \mathcal{Q}_n(d)} \sup_{\mu^n \in \Lambda^n} \left[ H_{\sigma(\bar{\pi})}(\mu^n) - \min_{\pi^* \in \mathcal{Q}_n(d)}   H_{\sigma(\pi^*)}(\mu^n)  \right].
\end{align}
This problem can be interpreted as the universal minimax counterpart of the problem presented in (\ref{eq_sec_pre_5b}).

\subsection{
Strong-Minimax Universality} 
\label{sub_sec:main_general_result}
From the analysis made on the two terms in (\ref{eq_main_proof_pre_1}), one could notice that everything reduces to the selection of the first-stage of the encoding process (the quantization). 
The following result formalizes this observation:
\begin{theorem}
\label{th_necessary_suffuficient_cond_universality}
	A necessary and sufficient condition for the existence of a strongly universal $D$-semifaithful code operating 
	at distortion $d>0$ for $\Lambda$ (Def. \ref{def_strong_minimax_universality}) is that there is a sequence of partitions $\left\{\pi_n, n \geq 1 \right\}$ satisfying the following:
	\begin{itemize}
		\item[i)] $\pi_n \in \mathcal{Q}_n(d)$ for all $n\geq 1$, (the fixed distortion requirement)
		\item[ii)]  
		$\lim_{n \rightarrow \infty} \frac{1}{n} R^+(\Lambda^n, \sigma(\pi_n)) =0$,  and 
		\item[iii)]  
		$\lim_{n \rightarrow \infty} \frac{1}{n}  \sup_{\mu^n \in  \Lambda^n} \left[ H_{\sigma({\pi_n})}(\mu^n) - \min_{\pi \in   \mathcal{Q}_n(d)}   H_{\sigma(\pi)}(\mu^n)  \right]=0$.
	\end{itemize}
\end{theorem}

From this result achieving strong minimax universality for $\Lambda$ at distortion $d$ requires meeting two important conditions: on the one hand,  that a universal quantizer can be found that approximates the best performance stated in (\ref{eq_sec_pre_5b}) as the block-length tends to infinity (the approximation criterion in iii)), and, on the other hand,  that the resulting information radius of the projected family grows sub-linearly with the block-length  (the complexity criterion in ii)).  This result captures the information radius condition known in the lossless universal source coding problem, but adds another component making the problem conceptually more difficult to address, which is the existence of a universal quantization for the family $\left\{\Lambda^n, n\geq 1 \right\}$ in the sense of condition iii). 

In this fixed-distortion  setting,  we could move to the extreme of asking for a zero distortion ($d=0$), where for any reasonable distortion, the quantizer $\phi_n$ needs to be the identity to meet the distortion criterion in i). In this context,  condition iii) is trivially met and minimax universality reduces to verifying the information radius condition of the un-projected family, i.e.,  
$R^+(\Lambda^n)  = \min_{v\in \mathcal{P}(\mathbb{X}^n)}  \sup_{\mu^n \in \Lambda^n}  D_{\sigma(\pi_{\phi})} (\mu^n \|v).$
Then, in the zero distortion regime, Theorem \ref{th_necessary_suffuficient_cond_universality}  recovers the necessary and sufficient condition known for lossless universal source coding \cite{csiszar_2004, gassiat_2018, boucheron_2009}.

In the next section,  we will use these conditions implicitly and explicitly to  study strong minimax universality  for the family of envelope distributions on infinite alphabets.

\subsection{Proof of Theorem \ref{th_necessary_suffuficient_cond_universality}}
\label{proof_th_necessary_suffuficient_cond_universality}
\begin{proof}
For the direct part, for any $n\geq 1$ and $d>0$,  let us consider a lossy code $\xi^*_n=(\phi^*_n, \mathcal{C}^*_n, \mathcal{D}^*_n)$ of length $n$ such that $\phi^*_n$ is determined from $\pi_n$, i.e. $\pi_{\phi^*_n}=\pi_n$.  From this,  $\xi_n$ is a $D$-semifaithful code operating at distortion $d$ from the assumption that $\pi_n \in \mathcal{Q}_n(d)$.\footnote{To achieve this, it is sufficient to have that $y^n \in {\phi^*_n}^{-1}(\left\{ y^n \right\})$ for any $y^n\in \mathcal{B}_n$. 
} For the second stage (the variable length encoder-decoder of $\mathcal{B}_n$), let us consider the pairs $( \mathcal{C}^*_n, \mathcal{D}^*_n)$  as a solution of 
the minimax problem presented in (\ref{eq_main_proof_pre_3}), i.e.,
$$\min_{(\mathcal{C}_n, \mathcal{D}_n)}  \sup_{\mu \in \Lambda_f} \left[ R({(\phi^*_n, \mathcal{C}_n, \mathcal{D}_n)}, \mu^n)  - \frac{H_{\sigma(\pi_{\phi^*_n})}(\mu^n)}{n}   \right].$$
Then we know from (\ref{eq_main_proof_pre_3}) that
\begin{align} \label{pr_th_necessary_suffuficient_cond_universality_1}
	\sup_{\mu \in \Lambda_f} \left[ R( \underbrace{\xi^*_n}_{(\phi^*_n, \mathcal{C}^*_n, \mathcal{D}^*_n)}, \mu^n)  - \frac{H_{\sigma(\pi_{\phi^*_n})}(\mu^n)}{n}   \right]   &\leq   \frac{1}{n} R^+(\Lambda_f^n, \sigma(\pi_{\phi^*_n})) +\frac{1}{n} \nonumber\\
	&=   \frac{1}{n} R^+(\Lambda_f^n, \sigma(\pi_n)) +\frac{1}{n}.
\end{align}
Using (\ref{eq_main_proof_pre_1}), it follows that 
\begin{align} \label{pr_th_necessary_suffuficient_cond_universality_2}
\sup_{\mu \in \Lambda_f} \left[ R(\xi^*_n, \mu^n) - \mathcal{R}_n(d,\mu^n)  \right]  &\leq  \sup_{\mu \in \Lambda_f} \left[ R(\xi^*_n, \mu^n)  - \frac{H_{\sigma(\pi_{\phi^*_n})}(\mu^n)}{n}   \right]  +  \sup_{\mu \in \Lambda_f}  \left[ \frac{H_{\sigma(\pi_{\phi^*_n})}(\mu^n)}{n}  - \mathcal{R}_n(d,\mu^n) \right] \nonumber\\
			&=\frac{1}{n}  \left(  R^+(\Lambda_f^n, \sigma(\pi_n)) +  \sup_{\mu^n \in  \Lambda^n_f} \left[ H_{\sigma({\pi_n})}(\mu^n) - \min_{\pi^* \in   \mathcal{Q}_n(d)}   H_{\sigma(\pi^*)}(\mu^n)  \right] + 1 \right), 
\end{align}
which concludes the proof from the assumptions on $\left\{\pi_n, n\geq 1 \right\}$.

For the other implication (i.e., the necessary condition), let us assume that there is a D-semifaithful coding scheme $\left\{\xi^*_n=(\phi^*_n, \mathcal{C}^*_n, \mathcal{D}^*_n), n\geq 1 \right\}$ operating at distortion $d>0$ such that
\begin{align} \label{pr_th_necessary_suffuficient_cond_universality_3}
	\lim_{n \rightarrow \infty} \sup_{\mu^n \in \Lambda^n}   \left[ R(\xi^*_n, \mu^n) - \mathcal{R}_n(d,\mu^n) \right] =0. 
\end{align}
From  (\ref{eq_sec_pre_0b}), we have that $\pi_{\phi^*_n} \in \mathcal{Q}_n(d)$ for all $n\geq 1$ (condition i)). Concerning the information radius, using (\ref{eq_main_proof_pre_3}) and (\ref{eq_main_proof_pre_3a}) it follows that:
\begin{align} \label{pr_th_necessary_suffuficient_cond_universality_4}
 \sup_{\mu^n \in \Lambda^n} \left[ R(\underbrace{(\phi^*_n, \mathcal{C}_n, \mathcal{D}_n)}_{\xi^*_n}, \mu^n)  - \frac{H_{\sigma(\pi_{\phi^*_n})}(\mu^n)}{n}   \right]  &\geq \min_{(\mathcal{C}_n, \mathcal{D}_n)}  \sup_{\mu^n \in \Lambda^n} \left[ R((\phi^*_n, \mathcal{C}_n, \mathcal{D}_n), \mu^n)  - \frac{H_{\sigma(\pi_{\phi^*_n})}(\mu^n)}{n}   \right] \nonumber\\
    &\geq  \frac{1}{n} R^+(\Lambda^n, \sigma(\pi_{\phi^*_n})).
\end{align}
Then using the decomposition of the average redundancy in (\ref{eq_main_proof_pre_1}), it follows that
\begin{align} \label{pr_th_necessary_suffuficient_cond_universality_5}
	\sup_{\mu^n \in \Lambda^n}  \left(  R(\xi^*_n, \mu^n) - \mathcal{R}_n(d,\mu^n) \right) \geq  \sup_{\mu^n \in \Lambda^n} \left[ R({\xi^*_n}, \mu^n)  - \frac{H_{\sigma(\pi_{\phi^*_n})}(\mu^n)}{n}   \right] \geq  \frac{1}{n} R^+(\Lambda_f^n, \sigma(\pi_{\phi^*_n})),
\end{align}
which proves that condition ii) is satisfied from (\ref{pr_th_necessary_suffuficient_cond_universality_3}). 
Using again (\ref{eq_main_proof_pre_1}), it follows that $\forall \mu^n \in \Lambda^n$
\begin{align} \label{pr_th_necessary_suffuficient_cond_universality_6}
	R(\xi^*_n, \mu^n) - \mathcal{R}_n(d,\mu^n) \geq  \frac{H_{\sigma(\pi_{\phi^*_n})}(\mu^n)}{n}  - \mathcal{R}_n(d,\mu^n).
\end{align}
Verifying condition iii) follows from (\ref{pr_th_necessary_suffuficient_cond_universality_3}) and the definition of $\mathcal{R}_n(d,\mu^n)$ in (\ref{eq_sec_pre_5b}).
\end{proof}

\section{Results For Envelope Families} 
\label{sec_main}
The results  for envelope distributions on $\infty$-alphabets are presented in this section.  
Let us first introduce some definitions that will be needed for the statement of results. We  begin introducing the family of models: 
\begin{definition}
\label{def_envelop}
Let $f:\mathbb{X} \longrightarrow \mathbb{R}^{+}$ be a non-negative function. 
We define the envelope family induced by $f$ as:
\begin{equation}\label{eq_sub_sec_UdSC_2}
	\Lambda_f \equiv \left\{ \mu \in  \mathcal{P}(\mathbb{X}): \mu(x)\leq f(x), \forall x  \in \mathbb{X}   \right\},
\end{equation}
where $(\mu(x))_{x\in \mathbb{X}}$ is a convenient short-hand notation for the probability mass function (pmf) of $\mu$.
\end{definition}

\begin{definition}  
Let $\mathcal{H}(\mathbb{X}) \subset \mathcal{P}(\mathbb{X})$ denote the set of all probabilities (source) with finite entropy in $\mathbb{X}$.
\end{definition}

In addition,  we need to introduce a notion of regularity for the distortion function.  We consider 
the Euclidean norm between  two points in $\mathbb{X}$ denoted by $ \left|i-j \right|$ for any $i,j \in \mathbb{X}$. 
With this, the blown-up ball of radius $\epsilon$ and centered at $i$  is denoted by $B_{\epsilon}(i) \equiv \left\{j\in \mathbb{X},  \left|i-j \right| < \epsilon \right\}$  for any $\epsilon>0$ and $i\in \mathbb{X}$. 

\begin{definition}\label{def_unbounded_consistent_dist}
An unbounded distortion function $\rho: \mathbb{X}\times \mathbb{X} \longrightarrow \mathbb{R}^+$
is said to be consistent with respect to the Euclidean norm if for any $K>0$, 
there exists $\epsilon>0$ such that for any $i\in \mathbb{X}$ if $j\notin B_{\epsilon}(i)$ then $\rho(i,j) \geq K$.
\end{definition}

\subsection{Main Results}
\begin{theorem} \label{main_th_part1}
Let $\Lambda_f \subset \mathcal{P}(\mathbb{X})$ be induced by a non-negative function $f$ and
$\rho$ be an unbounded distortion consistent with respect to the Euclidean norm (Def. \ref{def_unbounded_consistent_dist}). 
We have the following results:
\begin{enumerate}
	\item[i)] If $f\notin \ell_1(\mathbb{X})$,  then for any $d>0$ and any D-semifaithful coding scheme 
	$\left\{\xi_n,  n\geq 1 \right\}$ operating at distortion $d$: 
	\begin{equation*} 
		 \sup_{\mu \in \Lambda_f}   \left[ R(\xi_n, \mu^n) - \mathcal{R}_n(d,\mu^n)  \right]  =\infty, \ \forall  n\geq 1.
	\end{equation*}
	\item[ii)] If $f\in \ell_1(\mathbb{X})$,   then for any distortion $d>0$, there exists a D-semifaithful coding scheme 
	$\left\{\xi_n,  n\geq 1 \right\}$ operating at distortion $d$ --- with respect to $\left\{\rho_n, n\geq 1 \right\}$ --- that is weakly minimax universal, i.e., 
	\begin{equation*}
		\lim_{n \rightarrow \infty}   \left[ R(\xi_n, \mu^n) - \mathcal{R}_n(d,\mu^n) \right] =0,  
	\end{equation*}
	for any $\mu\in \Lambda_f \cap \mathcal{H}(\mathbb{X})$.
	\item[iii)] If $\sup_{\mu\in \Lambda_f}H(\mu)< \infty$, or, equivalently,  if $\Lambda_f\subset \mathcal{H}(\mathbb{X})$:\footnote{This condition implies that $f\in \ell_1(\mathbb{X})$.} then the same construction presented in ii) is strongly minimax universal, i.e., 
	\begin{equation*}
		\lim_{n \rightarrow \infty} \sup_{\mu \in \Lambda_f}   \left[ R(\xi_n, \mu^n) - \mathcal{R}_n(d,\mu^n) \right]  =0.  
	\end{equation*} 	
\end{enumerate}
\end{theorem}
The proofs are presented in Section \ref{sec_proof_main}.

Some remarks about Theorem \ref{main_th_part1}:

 {\bf 1: } The result in part i) implies that achieving strong-minimax universality is not feasible for the entire collection of stationary memoryless sources in $\infty$-alphabets.  This is a direct implication of this result using $f(i)=1$ for all $i\in \mathbb{X}$.
 
 {\bf 2: }
Interestingly, part i) matches the impossibility condition known for the lossless case in~\cite{boucheron_2009}. Therefore, in  the context of infinite alphabet sources, the non-zero distortion does not help making feasible the task of universal source coding as we move from the lossless to the lossy (fixed-distortion) setting of the variable length coding problem.

{\bf 3:} 
The argument used for the impossibility part relies on the proof of Theorem \ref{th_necessary_suffuficient_cond_universality}  and in particular on bounding from below the worse-case redundancy by the I-radius of $\Lambda_f$ projected over the cells induced by a $D$-semifaithful code (operating at distortion $d$). Then, the proof reduces to  show that this redefined I-radius (see (\ref{eq_main_proof_pre_3b})) is unbounded for any partition of $\mathbb{X}$ that belongs to $\mathcal{Q}_n(d)$ and for any $d>0$.

{\bf 4:} On the other hand assuming that $f\in \ell_1(\mathbb{X})$, the result in part ii) shows that there is a $D$-semifaithful scheme that achieves weak minimax universality for any $d>0$. This result is strengthened in part iii) showing that the same $D$-semifaithful construction 
is strong minimax universal provided that $\Lambda_f \subset \mathcal{H}(\mathbb{X})$. 

{\bf 5:} 
The constructive argument used for the proof of Theorem \ref{main_th_part1} (part iii) is based on a two-stage (lossy-lossless) scheme (see Figure \ref{fig1} in Section \ref{sec_proof_main}). The basic idea of this construction is to consider a specific two-stage lossy coding scheme.  In the first-stage of this scheme,  the problem is projected (loosely) to a finite alphabet task for which results for finite alphabet universal source coding are adopted (see Lemma \ref{lemma_finite_alphabet_unievrsal_construction} in Section \ref{sec_proof_main}). The second-stage, on the other hand,  is addressed  as a lossless source coding problem over a transformed infinite alphabet, where results from lossless universal source coding for envelope families are used  (see Lemma \ref{lemma_bound_inf_radius_envelop} in Section \ref{sec_proof_main}). 

{\bf 6:}
An important result used in the proof of Theorem \ref{main_th_part1} (part iii) is that the so called {\em envelope distribution} $\tilde{\mu}_f$ derived from $f$ by 
\begin{align*}
	\tilde{\mu}_f(x) \equiv \left\{ \begin{array}{ll}   f(x) & \textrm{if $x \geq \tau_f$}\\
			 					  1-\sum_{x\geq u_f} f(x) & \textrm{if $x=\tau_f-1$}\\
								  0 & \textrm{if $x < \tau_f-1$,}
 					  \end{array} 
					\right., 
\end{align*}
with $\tau_f \equiv  \min \left\{k\geq 1, \sum_{x\geq k} f(x)\leq 1 \right\}$, 
is the probability in $\Lambda_f$ that achieves maximum entropy under some mild considerations.  The formal statement of this result is presented in Lemma \ref{pro_maximun_entropy_tail_family} (in Sect \ref{sec_main_proof_achievability}). Therefore, the condition $\Lambda_f \subset \mathcal{H}(\mathbb{X})$ reduces to verify that $H(\tilde{\mu}_f)< \infty$
and, consequently, that the function $(f(x) \log 1/f(x))_{x\in \mathbb{X}}$ is summable.

{\bf 7:} Finally, Theorem \ref{main_th_part1} can be extended to the scenario of a bounded distortion if it is consistent with the Euclidean norm in the following sense:
\begin{definition}\label{def_bounded_dist}
A bounded distortion function $\rho: \mathbb{X}\times \mathbb{X} \longrightarrow [0,\rho_{max}]$, with 
$\rho_{max}>0$ ,  
is said to be consistent with respect to the Euclidean norm if for any $K\in (0,\rho_{max}]$, 
there is $\epsilon>0$ such that for any $i\in \mathbb{X}$ if $j\notin B_{\epsilon}(i)$ then $\rho(i,j) \geq K$.
\end{definition}
The statement of that result would be the same as the statement of Theorem \ref{main_th_part1} but restricting $d$ to the range  $(0,\rho_{max})$. The proof argument follows directly from the proof of Theorem \ref{main_th_part1}, consequently, both the statement and the proof are omitted. Finally, it is worth  noting that the {\em Hamming distance} satisfies Def.~\ref{def_bounded_dist} as many other regular distortions, e.g., $\rho_M(i,j) \equiv K \min \left\{\left|i - j \right|, M\right\} $ for any $K\in \mathbb{R}^+\setminus \left\{0 \right\}$ and $M>1$.

\subsection{Rate of Convergence}
The next result complements Theorem \ref{main_th_part1} by providing an upper bound on the rate of convergence for the worse-case overhead for the case of summable envelope families. 
\begin{theorem}\label{main_th_rate_overhead_envelope}
	Under the setting of Theorem \ref{main_th_part1}, 
	if $\Lambda_f\subset \mathcal{H}(\mathbb{X})$, and we add the condition that
	\begin{equation*}
		\lim \sup_{k \rightarrow  \infty} \frac{ \sum_{i\geq k} f(i) \log 1/ f(i) }{  \tilde{\mu}_f(\mathcal{T}_k) \log 1/ \tilde{\mu}_f(\mathcal{T}_k)} < \infty
	\end{equation*}
	with $\mathcal{T}_k \equiv \left\{k,k+1,\ldots \right\} \subset \mathbb{X}$,   
	then for any distortion $d>0$, there is a D-semifaithful coding scheme 
	$\left\{\xi^*_n,  n\geq 1 \right\}$ operating at distortion $d$  --- with respect to $\left\{\rho_n, n\geq 1 \right\}$ --- such that: 
		\begin{equation*} 
			\sup_{\mu \in \Lambda_f}  \left[ R(\xi^*_n, \mu^n) - \mathcal{R}_n(d,\mu^n) \right]  \leq C_0 \frac{u_f(n) \log n}{n} + C_1 \frac{\log n}{n} + C_2 \frac{1}{n}, 
		\end{equation*}
	where $C_0,C_1$ and $C_2$ are constants and 
	$$u_f(n) \equiv \min \left\{k\geq 1 \text{ such that } \tilde{\mu}_f(\mathcal{T}_{k+1})< 1/n \right\}.$$
\end{theorem}
The proof  is presented in Section \ref{sec_proof_main}.

This last result adds a regularity assumption on the way the tail component of the entropy of $\tilde{\mu}_f$ 
tends to zero, which is sufficient to obtain a rate of convergence for the worse-case overhead that is $O(u_f(n)\log (n)/n)$.
Importantly, it can be verified that polynomial envelope families (with $f_p(x)=1/x^p$ for some $p>1$) and 
exponential envelope families (with $f_p(x)=K e^{-\alpha x}$ with $K>0$ and $\alpha >0$)  satisfy the tail 
conditions stated in this result, and,  consequently, they are both strongly minimax universal. 
In fact, we have the following:
\begin{lemma}\label{prop_polinomial_envelop}
	Let us consider a polynomial function given by $(f_p(i))_{i\geq 1}=(1/i^p)_{i\geq 1}$. For any $p>1$ it follows that
	\begin{equation*}
		\lim \sup_{k \rightarrow  \infty} \frac{ \sum_{i\geq k} f_p(i) \log 1/ f_p(i) }{  \tilde{\mu}_{f_p}(\mathcal{T}_k) \log 1/ \tilde{\mu}_{f_p}(\mathcal{T}_k)} < \infty.
	\end{equation*}\vspace{1mm}
\end{lemma}
\begin{lemma}\label{prop_exponential_envelop}
	Let us consider an exponential function given by $(f_\alpha(i))_{i\geq 1}=(K e^{-\alpha i} ) _{i\geq 1}$.  For any $K>1$ and $\alpha>0$
	it follows that
	\begin{equation*}
		\lim \sup_{k \rightarrow  \infty} \frac{ \sum_{i\geq k} f_\alpha(i) \log (1/ f_\alpha(i))}{  \tilde{\mu}_{f_\alpha}(\mathcal{T}_k) \log (1/ \tilde{\mu}_{f_\alpha}(\mathcal{T}_k))} < \infty.
	\end{equation*}\vspace{1mm}
\end{lemma}
The proofs of these Lemmas are presented in Appendices \ref{proof_prop_polinomial_envelop} and  \ref{proof_prop_exponential_envelop}, respectively.

Finally, the sequence $(u_f(n))_{n\geq 1}$ was introduced by Bontemps {\em et al.} in \cite{bontemps_2014}  for the lossless source coding problem, where the same rate $O(u_f(n)\log (n)/n)$ was obtained for the redundancy of the best (lossless) universal scheme with $f\in \ell_1(\mathbb{X})$.

\section{Proofs of the Main Results of Section \ref{sec_main}}
\label{sec_proof_main}

\subsection{Theorem  \ref{main_th_part1} --- Part i): $f\notin \ell_1(\mathbb{X})$}
\label{sec_main_proof_imposibility}

\begin{proof}
Let us consider $d>0$ and arbitrary D-semifaithful coding scheme $\left\{\rho_n=(\phi_n, \mathcal{C}_n, \mathcal{D}_n), n\geq 1 \right\}$, such that 
$$\rho_n(x^n, \phi_n(x^n))\leq d,$$  
for all $n\geq 1$  and $x^n\in \mathbb{X}^n$. We denote by $\mathcal{B}_n=\left\{ \phi_n(x^n), x^n \in \mathbb{X}^n  \right\}$ the range of $\phi_n$ and by $\pi_{\phi_n}$ the partition induced by $\phi_n$ (see Eq.(\ref{eq_sec_pre_0c})). 
From the decomposition in (\ref{eq_main_proof_pre_1}), for any $\mu^n \in \Lambda^n_f$
\begin{align}\label{eq_main_proof_pre_5}
R(\xi_n, \mu^n) - \mathcal{R}_n(d,\mu^n) \geq \left[ R(\xi_n, \mu^n)  - \frac{H_{\sigma(\pi_{\phi_n})}(\mu^n)}{n}   \right].
\end{align}
From (\ref{eq_main_proof_pre_5}) and the 
analysis presented in Sec.\ref{sec_proof_main_pre_information_radius},  
the worse-case overhead  over $\Lambda_f$ is bounded by 
\begin{align}\label{eq_main_proof_pre_6}
\sup_{\mu^n \in \Lambda^n_f}   R(\xi_n, \mu^n) - \mathcal{R}_n(d,\mu^n) 
				&\geq  \frac{1}{n}  \sup_{\mu \in \Lambda_f} D(v_{\mu^n}\|v_{\mathcal{C}_n}) \nonumber\\
				& \geq \frac{1}{n}  \min_{v\in \mathcal{P}(\mathbb{X}^n)}  \sup_{\mu^n \in \Lambda^n_f}  D_{\sigma(\pi_{\phi})} (\mu^n \|v)\nonumber\\
				& = \frac{1}{n} R^+(\Lambda_f^n, \sigma(\pi_{\phi_n})),
\end{align}
where $R^+(\Lambda_f^n, \sigma(\pi_{\phi_n}))$ is the information radius of the family $\Lambda_f^n$
restricted to the sub-sigma field induced by $\pi_{\phi_n}$.

The rest of the proof  will show that $R^+(\Lambda_f^n, \sigma(\pi_{\phi_n}))=\infty$, for any $n\geq 1$.
Using that $f\notin \ell_1(\mathbb{X})$, i.e., $\sum_{x\in \mathbb{X}} f(x)= \infty$, 
we  follow  ideas used in  lossless coding  \cite{boucheron_2009}, to construct 
a countable collection of distributions $\tilde{\Lambda} = \left\{ \tilde{\mu}_j, j\in \mathcal{J} \right\} \subset \Lambda_f$
with $\left| \mathcal{J} \right|  =  \infty$, where $\mathcal{A}_j=\text{supp}(\tilde{\mu}_j)$  is such that $\left| \mathcal{A}_j\right| <\infty$ 
and for any $i,j\in \mathcal{J}$ $i \neq j$ $\mathcal{A}_i\cap \mathcal{A}_j = \emptyset$. 
Then we can consider the $n$-fold family 
$\tilde{\Lambda}^n=\left\{ \tilde{\mu}^n_j, j\in \mathcal{J} \right\} $ where $support(\tilde{\mu}^n_j)=\mathcal{A}^n_j=\mathcal{A}_j\times ...\times A_j\subset \mathbb{X}^n$. Using the consistency of $\rho_n$ with respect to the Euclidean norm (in Def.\ref{def_unbounded_consistent_dist}), it follows that  to achieve the  distortion 
criterion in (\ref{eq_sec_pre_0b}),  it is necessary that the range of  $\phi_n$ has an infinite number of prototypes (i.e., $\left|\mathcal{B}_n \right|=\infty$), otherwise, it is simple to verify that $\sup_{x^n\in \mathbb{X}^n}\min_{y^n\in \mathcal{B}_n} \rho_n(x^n,y^n)>d$.

For any $j\in \mathcal{J}$, let us consider a covering of the support of $\tilde{\mu}^n_j$ by cells of $\pi_{\phi_n}$ by
\begin{align}\label{eq_main_proof_pre_7}
	\mathcal{C}({\mathcal{A}}^n_j) \equiv \bigcup_{\mathcal{B}\in \pi_n({A}^n_j)} \mathcal{B}, 
\end{align}
where $\pi_n({\mathcal{A}}^n_j) \equiv  \left\{\mathcal{B}\in \pi_{\phi_n}, {\mathcal{A}}^n_j \cap \mathcal{B} \neq \emptyset \right\}$. 
At this point,  we  can show that $\left| \mathcal{C}({\mathcal{A}}^n_j) \right|< \infty$, $\forall j\in \mathcal{J}$. This follows from the construction of $\left\{ {\mathcal{A}}^n_j , j \in \mathcal{J} \right\}$ 
and the observation that  any $\mathcal{B}$ in  $\pi_{\phi_n}$  is a finite set
from the hypothesis that $\pi_{\phi_n}\in \mathcal{Q}_n(d)$ and the consistency assumption on  $\rho_n$ (Def. \ref{def_unbounded_consistent_dist}).
Therefore,  we get  that $\mathcal{C}({\mathcal{A}}^n_j)$ in (\ref{eq_main_proof_pre_7}) is a finite set for any $j$. 

Let us consider a countably infinite sub-collection of disjoint sets in $\left\{ \mathcal{C}({\mathcal{A}}^n_j), j \in \mathcal{J}  \right\}$  by the following approach:
\begin{align}\label{eq_main_proof_pre_8}
	{j}_1 &\equiv  1,\nonumber\\
	{j}_2 &\equiv \min \left\{j>j_1, \text{ such that }  \mathcal{C}({\mathcal{A}}^n_j) \cap \mathcal{C}({\mathcal{A}}^n_{j_1})=\emptyset \right\}, \nonumber\\
	\ldots  &	\nonumber\\
	{j}_k &\equiv \min \left\{j>j_{k-1}, \text{ such that }  \mathcal{C}({\mathcal{A}}^n_j) \cap \bigcup_{l=1}^{k-1} \mathcal{C}({\mathcal{A}}^n_{j_l})=\emptyset \right\} \ldots .
\end{align}
For any finite $k$, the solution in (\ref{eq_main_proof_pre_8}) is guaranteed to be achieved with a finite integer; then, we have 
an infinite new collection of probabilities $\hat{\Lambda}^n \equiv  \left\{\tilde{\mu}^n_{j_k}, k\geq 1 \right\} \subset \tilde{\Lambda}^n \subset \Lambda^n_f$. 
Based on the  construction of $\tilde{\mu}^n_{j_k}$, the family  $\hat{\Lambda}^n$ is composed of a collection of probabilities with disjoint support  in $\mathbb{X}^n$.   Then, we consider the following partition of $\mathbb{X}^n$
\begin{align}\label{eq_main_proof_pre_9}
	\eta_n \equiv  \left\{ \mathcal{C}({\mathcal{A}}^n_{j_k}), k\geq 1 \right\} \cup  \left( \mathbb{X}^n \setminus  \bigcup_{k=1}^{\infty} \mathcal{C}({\mathcal{A}}^n_{j_k}) \right), 
\end{align}
where it is clear that $\sigma(\eta_n) \subset \sigma(\pi_{\phi_n})$ 
and for any $\mu,  v \in \mathcal{P}(\mathbb{X}^n)$, 
$D_{\sigma(\eta_n) }(\mu \| v) \leq D_{\sigma(\pi_{\phi_n}) }(\mu \| v)$.  
The important point here is that  $\hat{\Lambda}^n$ 
contains an infinite set of distributions with disjoint support when restricted to the cells of $\eta_n$ and,  thus,  from the known connection 
between information radius and channel capacity  \cite{csiszar_2004}, the following can be obtained: 
\begin{lemma} \label{prop_infinite_information_radius}
	$R^+(\hat{\Lambda}^n, \sigma(\eta_n))=\infty$.
\end{lemma}
The proof of Lemma \ref{prop_infinite_information_radius} is presented in Appendix \ref{proof_prop_infinite_information_radius}.

Therefore, we have that
\begin{align}\label{eq_main_proof_pre_10}
	R^+(\Lambda^n_f, \sigma(\pi_{\phi_n})) \geq  R^+(\hat{\Lambda}^n, \sigma(\pi_{\phi_n})) \geq R^+(\hat{\Lambda}^n, \sigma(\eta_n))=\infty, 
\end{align}
from the fact that by construction $\hat{\Lambda}^n \subset \Lambda^n_f$ and $\sigma(\eta_n) \subset \sigma(\pi_{\phi_n})$. 
Finally (\ref{eq_main_proof_pre_10}) and  (\ref{eq_main_proof_pre_6}) prove the impossibility part (Theorem \ref{main_th_part1} i)).\footnote{Alternatively, this result can be derived from (\ref{eq_main_proof_pre_10}) and Theorem \ref{th_necessary_suffuficient_cond_universality}.}
\end{proof}

\subsection{Theorem  \ref{main_th_part1} --- Part iii): $f\in \ell_1(\mathbb{X})$ and $\Lambda_f\subset \mathcal{H}(\mathbb{X})$}
\label{sec_main_proof_achievability}
To organize the proof of this part,  let us first introduce preliminary results and definitions that will be used in the main argument. 
\begin{definition}\label{def_envelop_distribution}
The distribution induced by the tail function $f$ is given by: 
\begin{align}\label{eq_main_proof_pre_11}
	\tilde{\mu}_f(x) \equiv \left\{ \begin{array}{ll}   f(x) & \textrm{if $x \geq \tau_f$}\\
			 					  1-\sum_{x\geq u_f} f(x) & \textrm{if $x=\tau_f-1$}\\
								  0 & \textrm{if $x < \tau_f-1$,}
 					  \end{array} 
					\right.
\end{align}
where $\tau_f \equiv  \min \left\{k\geq 1, \sum_{x\geq k} f(x)\leq 1 \right\}$. 
\end{definition}
Note that by construction, we have that $\tilde{\mu}_f(x)\in \Lambda_f$ and $\tau_f<\infty$ from the hypothesis that  $f\in \ell_1(\mathbb{X})$.
%

Let us consider the finite set $\Gamma_k= \left\{1,.., k \right\}$ for any $k\geq 1$.  Then we have the following 
result for finite alphabet sources:
\begin{lemma} \label{lemma_finite_alphabet_unievrsal_construction}
	For any $n \geq 1$,  $k\geq 1$, distortion $d>0$ and $\epsilon>0$, there is a $D$-semifaithful code $\xi^{*k}_n=(\phi^{*k}_n,\mathcal{C}^{*k}_n, \mathcal{D}^{*k}_n)$ 	on 
	$\Gamma_{k+1}$,  that operates at distortion $d>0$ (w.r.t. $\tilde{\rho}_n$) and verifies that 
	\begin{equation*}
		 \sup_{v\in \mathcal{P}(\Gamma_{k+1})} \left[ \frac{1}{n} \mathbb{E}_{Y^n\sim v^n}  \left\{  \mathcal{L}(\mathcal{C}^{*k}_n(\phi^{*k}_n(Y^n))) \right\} -   \mathcal{R}_n(d,v^n) \right]   \leq  \frac{k \log(n+1)}{n} + \epsilon, 
	\end{equation*}
	where $\mathcal{P}(\Gamma_{k+1})$ is the collection of probabilities on $\Gamma_{k+1}$ (i.e., the simplex of dimension $k$).
\end{lemma}
The proof of Lemma \ref{lemma_finite_alphabet_unievrsal_construction} is presented in 
Appendix \ref{proof_lemma_finite_alphabet_unievrsal_construction}.

For envelope families on infinite alphabets, we have the following remarkable result from Bontemp {\em et al.} \cite{bontemps_2014}: 
\begin{lemma}\label{lemma_bound_inf_radius_envelop} \cite[Prop. 5]{bontemps_2014}
If $f\in \ell_1(\mathbb{X})$, then for any $n\geq 1$
\begin{align}\label{eq_main_proof_pre_33}
	(1+ o(1))\frac{u_f(n)-1}{4} \log n \leq  R^+(\Lambda_f^n) \leq 2+ \log e + \frac{u_f(n)-1}{2} \log n, 
\end{align}
where 
\begin{equation}\label{eq_main_proof_pre_33b}
	u_f(n) = \min \left\{k\geq 1 \text{ such that } \tilde{\mu}_f(\mathcal{T}_{k+1})< 1/n \right\}. 
\end{equation}
\end{lemma}

Finally, let us consider a tail partition of $\mathbb{X}$ given by $\tilde{\pi}_k \equiv \left\{ \Gamma_k, \left\{ k+1\right\}, \left\{ k+2\right\}, \ldots \right\}$
for any $k\geq 1$. The next result shows that the tail distribution $\tilde{\mu}_f$ (in Def. \ref{eq_main_proof_pre_11})
achieves  maximum entropy over the envelope family in the following sense:  
\begin{lemma} \label{pro_maximun_entropy_tail_family}
If $H(\tilde{\mu}_f)<\infty$, it follows that eventually in $k$ (i.e., for a sufficiently large $k$),
\begin{equation*}
	\sup_{\mu\in \Lambda_f }H_{\sigma(\tilde{\pi}_k)}(\mu)= H_{\sigma(\tilde{\pi}_k)}(\tilde{\mu}_f) < \infty.
\end{equation*}
Otherwise,  if $H(\tilde{\mu}_f) = \infty$, then  $\sup_{\mu\in \Lambda_f }H_{\sigma(\tilde{\pi}_k)}(\mu) =H_{\sigma(\tilde{\pi}_k)}(\tilde{\mu}_f) = \infty$ for any $k\geq 1$.
\end{lemma}
The proof of Lemma \ref{pro_maximun_entropy_tail_family} is presented in Appendix \ref{proof_pro_maximun_entropy_tail_family}.

\begin{proof}
The basic idea  of the proof is to decompose the alphabet $\mathbb{X}$ into two segments and use 
a two-stage scheme.   More precisely, let us consider the following mapping 
$S_k:\mathbb{X} \longrightarrow \Gamma_{k+1} = \left\{1,..,k+1 \right\}$ where
\begin{align}\label{eq_main_proof_pre_12}
	S_k(x) \equiv \left\{ \begin{array}{ll}   x & \textrm{if $x \in \Gamma_k = \left\{1,..,k \right\}$}\\
							 k+1 & \textrm{if $x> k$}
 					  \end{array} 
					\right.
\end{align}
Applying this lossy mapping (letter by letter) to the source $X^n$, we create a truncated version of it: 
\begin{align}\label{eq_main_proof_pre_13}
	Y^n_1(k)\equiv S_k(X^n) \equiv (S_k(X_1),\ldots, S_k(X_n))\in \Gamma_{k+1}^n.
\end{align}
To retain the information lost from $X^n$ in $Y^n_1(k)$, the following complementary mapping is used:
\begin{align}\label{eq_main_proof_pre_14}
	O_k(x) \equiv \left\{ \begin{array}{ll}   	1 & \textrm{if $x \in \Gamma_k$}\\
							 	x & \textrm{if $x> k$}
 					 	 	\end{array} 
							\right. \in  \left\{1 \right\} \cup \Gamma_k^c, 
\end{align}
which induces
\begin{align}\label{eq_main_proof_pre_15}
	Z^n_1(k)\equiv O_k(X^n) \equiv (O_k(X_1),\ldots, O_k(X_n))\in  (\left\{1 \right\} \cup \Gamma_k^c)^n.
\end{align}

It is clear that for any $k\geq 1$, $Y^n_1(k)$ and $Z^n_1(k)$ recover $X^n$ with no loss.
In this context,  we propose a two-stage strategy where $Y^n_1(k)$ (a finite alphabet stationary memoryless source)
is encoded with a $D$-semifaithful code (operating at distortion $d>0$) and $Z^n_1(k)$ (an infinite alphabet stationary memoryless 
source) is encoded losslessly  using a variable-length code.  
Let us consider a distortion $d>0$  and a   $D$-semifaithful triplet $\xi^k_n=(\phi^k_n,\mathcal{C}^k_n, \mathcal{D}^k_n)$ for the source $Y^n_1(k)$ on the alphabet $\Gamma_{k+1}$,   operating at distortion $d>0$  with respect to a distortion $\tilde{\rho}$ on $\Gamma_{k+1}\times \Gamma_{k+1}$,  where we  
assume that $\tilde{\rho}$ coincides with $\rho$ on $\Gamma_k \times \Gamma_k$ (the non-truncated symbols, see Eq. (\ref{eq_main_proof_pre_12})).
This means that for all $y^n\in \Gamma_{k+1}^n$
\begin{align}\label{eq_main_proof_pre_16}
	\tilde{\rho}_n (y^n, \phi^k_n(y^n)) \leq d. 
\end{align}
On the other hand, we can consider a lossless variable-length 
encoder-decoder pair $(\tilde{\mathcal{C}}_n^k, \tilde{\mathcal{D}}^k_n)$ for 
the source $Z^n_1(k)$, where  $\tilde{\mathcal{C}}^k_n: (\left\{1 \right\} \cup \Gamma_k^c)^n  \longrightarrow \left\{0,1 \right\}^*$ and $\tilde{\mathcal{D}}^k_n: \left\{0,1 \right\}^* \longrightarrow  (\left\{1 \right\} \cup \Gamma_k^c)^n$. Then,  given an 
input  $x^n\in \mathbb{X}^n$ the final output (after decoding) of this two-stage approach is 
\begin{align}\label{eq_main_proof_pre_17}
	(\hat{y}^n,z^n )=(\phi^k_n(S_k(x^n)), O_k(x^n))\in (\Gamma_{k+1})^n  \times  (\left\{1 \right\} \cup \Gamma_k^c)^n.
\end{align}
Finally,  we recover $\hat{x}^n$ from $(\hat{y}^n,z^n )$ by the following letter-by-letter mapping 
$$\hat{x}^n = (\Psi_k(\hat{y}_1,z_1), \ldots, \Psi_k(\hat{y}_n,z_n)) \in \mathbb{X}^n,$$
where
\begin{align}\label{eq_main_proof_pre_18}
	\Psi_k(\hat{y}_i,z_i) \equiv \left\{ \begin{array}{ll}   	z_i & \textrm{if $z_i \in \Gamma_k^c$}\\
							 				\hat{y}_i & \textrm{if $z_1=1$}
 					 	 					\end{array} 
							\right.\in \mathbb{X}.
\end{align}
Then, using the condition imposed on $\tilde{\rho}$, it follows that 
\begin{align}\label{eq_main_proof_pre_19}
	\rho_n(x^n, \hat{x}^n) \leq \tilde{\rho}_n(y^n, \hat{y}^n) \leq d, 
\end{align}
where $y^n=S_k(x^n)$ and $\hat{y}^n$ is defined in (\ref{eq_main_proof_pre_17}).
The first inequality in (\ref{eq_main_proof_pre_19}) is verified in Appendix \ref{proof_prop_distortion_ineq}, and the second follows from the 
fact that $\xi^k_n$ is a $D$-semifaithful code with respect to $\tilde{\rho}$. 
Therefore, this two-stage strategy produces a $D$-semifaithful  code in $\mathbb{X}^n$
with respect to $\rho$. The encoding-decoding process is illustrated in Figure \ref{fig1}.
\begin{figure}
\centering
\includegraphics[width=0.95\textwidth]{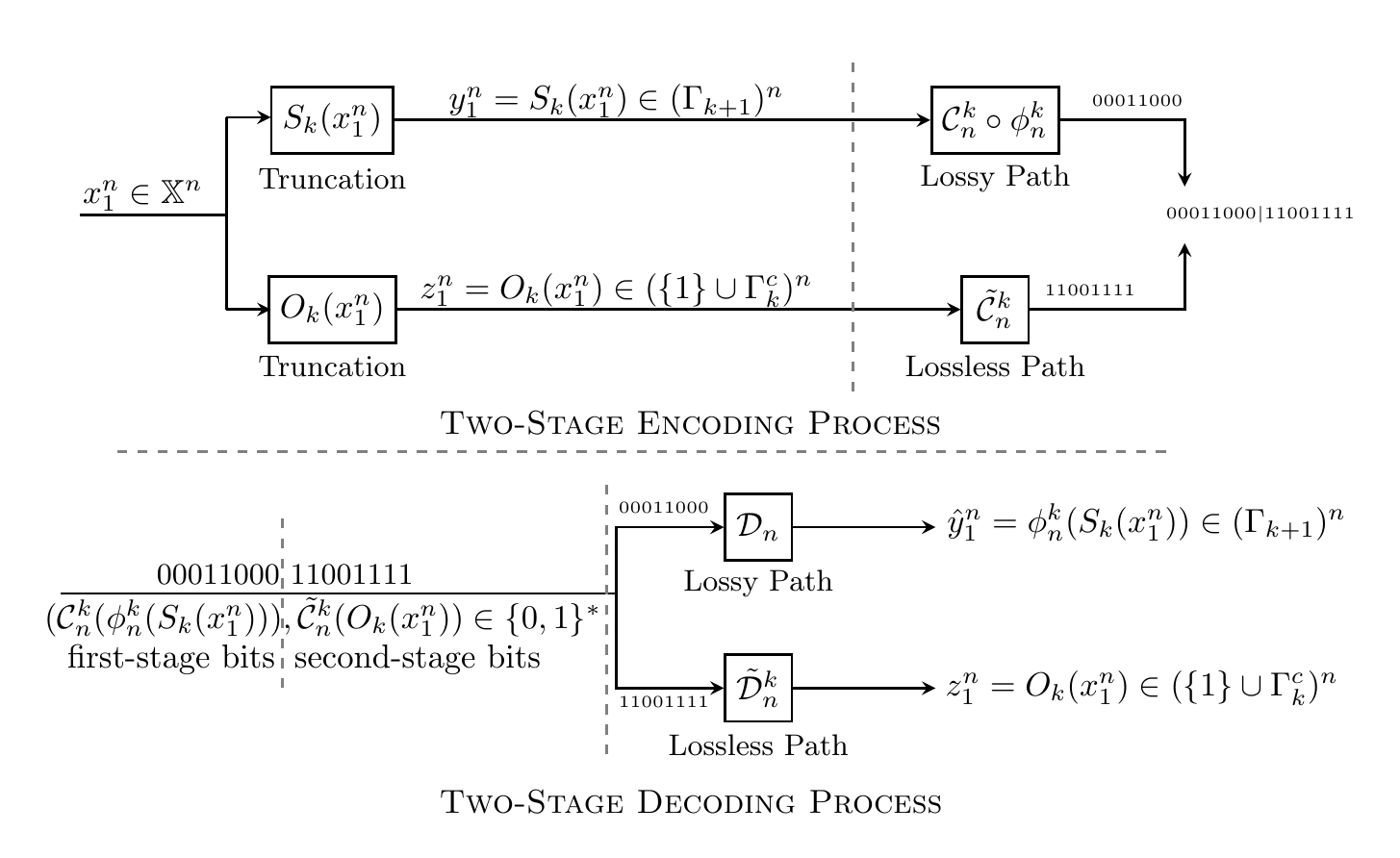}
\caption{Illustration of the two-stage scheme used in the achievability argument of Theorem \ref{main_th_part1} ($f\in \ell_1(\mathbb{X})$).}
\label{fig1}
\end{figure}

 On the other hand, the  length of this two-stage mapping (in bits per sample) that we denote by $\mathcal{T}^k_n$
 is given by: 
\begin{align}\label{eq_main_proof_pre_20}
	\frac{1}{n} \mathcal{L}(\mathcal{T}^k_n(x^n)) = \frac{1}{n} \left[  \mathcal{L}(\mathcal{C}^k_n(\phi^k_n(S_k(x^n))))  + \mathcal{L} (\tilde{\mathcal{C}}^k_n( O_k(x^n))) \right].
\end{align}
Then if $X^n \sim \mu^n$, the average length is given by: 
\begin{align}\label{eq_main_proof_pre_21}
	\frac{1}{n} \mathbb{E}_{X^n} \left\{ \mathcal{L}(\mathcal{T}^k_n(X^n)) \right\} = \frac{1}{n}  \underbrace{\mathbb{E}_{Y^n} \left\{  \mathcal{L}(\mathcal{C}^k_n(\phi^k_n(Y^n)))  \right\}}_{\text{first-stage bits}}  + \frac{1}{n}  \underbrace{\mathbb{E}_{Z^n}  \left\{ \mathcal{L}(\tilde{\mathcal{C}}^k_n(Z^n)) \right\}}_{\text{second-stage bits}},
\end{align}
where $Y^n=S_k(X^n)$ and $Z^n=O_k(X^n)$.

\subsubsection{Analysis of the first-stage bits in (\ref{eq_main_proof_pre_21})}
For the first term on the RHS of (\ref{eq_main_proof_pre_21}), it will be useful to consider the 
following truncated distortion $\rho^k$ on $\mathbb{X}\times \mathbb{X}$,  
\begin{align}\label{eq_main_proof_pre_22}
	\rho^k(x,\bar{x}) \equiv \left\{ \begin{array}{ll}   	\rho(x,\bar{x}) & \textrm{if $x,\bar{x} \in \Gamma_k$}\\
							 			0 & \textrm{if $x,\bar{x} \notin \Gamma_k$}\\
										\min_{\tilde{x}>k} \rho(x,\tilde{x}) & \textrm{if ${x} \in \Gamma_k$ and $\bar{x} \notin \Gamma_k$}\\
										\min_{\tilde{x}>k} \rho(\tilde{x},\bar{x}) & \textrm{if ${x} \notin \Gamma_k$ and $\bar{x} \in \Gamma_k$}
 	\end{array} 
	\right.,\ \forall x,\bar{x} \in \mathbb{X}. 
\end{align}
to specify  $\tilde{\rho}$ in $\Gamma_{k+1}\times \Gamma_{k+1}$,  used in the first-stage of the construction. 
It follows  that  $\rho^k(x,\bar{x})\leq \rho(x,\bar{x})$ and $\rho^k(x,\bar{x})=\rho^k(S_k(x),S_k(\bar{x}))$. Consequently,  we have that $\rho_n^k(x^n,\bar{x}^n)=\rho_n^k(S_k(x^n),S_k(\bar{x}^n))$
for any $x^n$ and $\bar{x}^n$ in $\mathbb{X}^n$. 
For the rest of the argument, we fix $\tilde{\rho}_n(y^n, \bar{y}^n)$ to be $\rho^k_n(y^n, \bar{y}^n)$ for any $y^n, \bar{y}^n\in \Gamma_{k+1}\times \Gamma_{k+1}$.
With this, let us introduce the counterpart of $\mathcal{R}_n(d,\mu_n)$ in (\ref{eq_sec_pre_5b}) but using instead 
the induced distortion $\rho^k_n$, i.e., 
\begin{align}\label{eq_main_proof_pre_22b}
\mathcal{R}^k_n(d,\mu^n) \equiv \min_{\pi \in \mathcal{Q}^k_n(d)}  \frac{H_{\sigma(\pi)}(\mu^n)}{n},
\end{align} 
where $\mathcal{Q}^k_n(d)$ is the collection of partitions of $\mathbb{X}^n$ such that any $\pi\in \mathcal{Q}^k_n(d)$ satisfies that $\forall A\in \pi$, $\exists y^n \in A$ such that $\sup_{x^n\in A}\rho^k_n(x^n,y^n) \leq d$. Then from the definition in (\ref{eq_sec_pre_5b}), 
we have that:
\begin{align}\label{eq_main_proof_pre_23}
	\mathcal{R}^k_n(d,\mu^n) \leq \mathcal{R}_n(d,\mu^n), 
\end{align}
for any $d>0$, any $n\geq 1$, any $k\geq 1$ and any $\mu \in  \mathcal{P}(\mathbb{X})$.

On the other hand, if we consider the distribution of $Y^n= S_k(X^n)\in \Gamma_{k+1}$ (assuming that $X^n\sim \mu^n$ for some marginal $\mu\in \mathcal{P}(\mathbb{X})$) 
and in particular  its marginal distribution $v_{\mu}$ in $\mathcal{P}(\Gamma_{k+1})$,  we can rely on the operational finite-length rate-distortion function $\mathcal{R}_n(d,v^n_\mu)$ in (\ref{eq_sec_pre_5b}). Using the fact that $\tilde{\rho}(S_k(x),S_k(x))=\rho^k(x,\bar{x})$, it is simple to show that
\begin{align}\label{eq_main_proof_pre_24}
	\mathcal{R}_n(d,v^n_\mu) = \mathcal{R}^k_n(d,\mu^n), 
\end{align}
for any $d>0$, any $n\geq 1$, any $k\geq 1$ and any $\mu \in  \mathcal{P}(\mathbb{X})$.

Finally, for any $D$-semifaithful code $\xi^k_n=(\phi^k_n,\mathcal{C}^k_n, \mathcal{D}^k_n)$  for $Y^n$ operating at distortion $d>0$ w.r.t. $\tilde{\rho}_n$, we have from (\ref{eq_main_proof_pre_24})  that 
\begin{align}\label{eq_main_proof_pre_25}
	\frac{1}{n} \mathbb{E}_{Y^n\sim v^n_\mu}  \left\{  \mathcal{L}(\mathcal{C}^k_n(\phi^k_n(Y^n))) \right\} \geq  \mathcal{R}_n(d,v^n_\mu) = \mathcal{R}^k_n(d,\mu^n).
\end{align}
At this  point,  we can use the result in Lemma \ref{lemma_finite_alphabet_unievrsal_construction}  for  finite alphabet sources. In particular, 
%
from Lemma \ref{lemma_finite_alphabet_unievrsal_construction} (choosing $\epsilon= 1/n$) and the expressions in (\ref{eq_main_proof_pre_25}) and (\ref{eq_main_proof_pre_24}),  we have that for any $n\geq 1$, $k\geq 1$ and distortion $d>0$, there is a $D$-semifaithful code $\xi^{*k}_n$ for the first-stage such that
\begin{align}\label{eq_main_proof_pre_26}
 \sup_{\mu \in \Lambda_f}  \left[  \frac{1}{n} \mathbb{E}_{X^n\sim \mu^n}   \left\{  \mathcal{L}(\mathcal{C}^{*k}_n(\phi^{*k}_n(S_k(X^n)))) \right\} - \mathcal{R}^k_n(d,\mu^n)  \right]   \leq  \frac{k \log(n+1)}{n} + \frac{1}{n}.  
\end{align}

\subsubsection{Analysis of the second-stage bits in (\ref{eq_main_proof_pre_21})}
Considering the second term in the RHD of (\ref{eq_main_proof_pre_21}),  let $m_\mu$ 
be the distribution of $Z_i=O_k(X_i)$ induced by $\mu$, then we have that  
\begin{align}\label{eq_main_proof_pre_27}
	\frac{1}{n} \mathbb{E}_{Z^n\sim m_\mu^n}  \left\{ \mathcal{L}(\tilde{\mathcal{C}}^k_n(Z^n)) \right\} \geq H(m_\mu), 
\end{align}
because $\tilde{\mathcal{C}}^k_n$ is a variable length (prefix-free) lossless encoder of $Z^n$ \cite{cover_2006}. Furthermore, it is well understood that the redundancy of $\tilde{\mathcal{C}}^k_n$ is equal to (up to a discrepancy of $O(1/n)$)
\begin{align}\label{eq_main_proof_pre_28}
	\frac{1}{n}  \left[  \mathbb{E}_{Z^n\sim m_\mu^n}  \left\{ \mathcal{L}(\tilde{\mathcal{C}}^k_n(Z^n)) \right\} - H(m_\mu^n)\right]  \approx \frac{1}{n} D(m^n_\mu \| m_{\tilde{\mathcal{C}}^k_n}),
\end{align}
where $m_{\tilde{\mathcal{C}}^k_n}\in \mathcal{P}((\left\{1 \right\} \cup \Gamma^c_k)^n)$ is the distribution associated with the prefix-free code $\tilde{\mathcal{C}}^k_n$ \cite{cover_2006,csiszar_2004}. From this observation, the criterion for designing the second-stage in the context of universal source coding  reduces to solving the following  problem\footnote{Using the correspondence between prefix-free codes and perfect (dyadic) distributions.}: 
\begin{align}\label{eq_main_proof_pre_29}
	R^{+}(\tilde{\Lambda}^n_f,k) \equiv \min_{m\in \mathcal{P}((\left\{1 \right\} \cup \Gamma^c_k)^n)} \sup_{\mu \in \Lambda_f} D(m^n_\mu\|m),
\end{align}
which is the information radius of the projected family $\tilde{\Lambda}^n_f \equiv \left\{m^n_\mu, \mu \in \Lambda_f \right\}$. 
In particular,  associated to the solution of (\ref{eq_main_proof_pre_29}) \cite{csiszar_2004}, there is a lossless code $\tilde{\mathcal{C}}^{*k}_n$ such that
\begin{align}\label{eq_main_proof_pre_30}
	\frac{R^{+}(\tilde{\Lambda}^n_f,k)}{n} \leq  \sup_{\mu \in \Lambda_f}    \left[ \frac{1}{n}  \mathbb{E}_{Z^n\sim m_\mu^n}  \left\{ \mathcal{L}(\tilde{\mathcal{C}}^{*k}_n(Z^n)) \right\} - H(m_\mu)\right] \leq  \frac{R^{+}(\tilde{\Lambda}^n_f,k)+1}{n}.
\end{align}
Importantly, using the information radius object introduced in (\ref{eq_main_proof_pre_3}),  it is simple to check that
\begin{align}\label{eq_main_proof_pre_31}
	R^{+}(\tilde{\Lambda}^n_f, k)  =  R^+(\Lambda_f, \sigma(\tilde{\pi}^{\times n}_k)) = \min_{m\in \mathcal{P}(\mathbb{X}^n)} \sup_{\mu \in \Lambda_f} D_{\sigma(\tilde{\pi}^{\times n}_k)}(\mu^n\|m),
\end{align}
where 
$$\tilde{\pi}^{\times n}_k \equiv \left\{ \Gamma_k, \left\{ k+1\right\}, \left\{ k+2\right\}, \ldots \right\} \times \left\{ \Gamma_k, \left\{ k+1\right\}, \left\{ k+2\right\}, \ldots \right\} \times ..\times .\left\{ \Gamma_k, \left\{ k+1\right\}, \left\{ k+2\right\}, \ldots \right\}$$ 
denotes the partition of $\mathbb{X}^n$ induced by the lossy mapping $(O_k(),O_k(),..,O_k()):\mathbb{X}^n \longrightarrow (\left\{1 \right\} \cup \Gamma_k)^n$. Then from (\ref{eq_main_proof_pre_3c}) and (\ref{eq_main_proof_pre_31})
\begin{align}\label{eq_main_proof_pre_32}
	R^{+}(\tilde{\Lambda}^n_f, k)  \leq R^+(\Lambda_f^n) \equiv  \min_{m\in \mathcal{P}(\mathbb{X}^n)} \sup_{\mu \in \Lambda_f} D(\mu^n\|m),
\end{align}
the last expression being the information radius of the unconstrained family $\Lambda_f^n$ \cite{csiszar_2004}.
%
The result by Bontemp et al. \cite{bontemps_2014}  (stated in Lemma \ref{lemma_bound_inf_radius_envelop}) 
for summable envelope families comes in handy here.  
%
In fact, integrating Lemma \ref{lemma_bound_inf_radius_envelop} 
in (\ref{eq_main_proof_pre_30}), for any $k\geq 1$ and $n\geq 1$,  there exists a variable-length 
code $\tilde{\mathcal{C}}^{*k}_n:(\left\{1 \right\} \cup \Gamma_k^c)^n \longrightarrow \left\{ 0,1\right\}^*$
satisfying that: 
\begin{align}\label{eq_main_proof_pre_34}
	 \sup_{\mu \in \Lambda_f}    \left[ \frac{1}{n} \mathbb{E}_{Z^n\sim m^n_\mu}  \left\{ \mathcal{L}(\tilde{\mathcal{C}}^{*k}_n(Z^n)) \right\} - H(m_\mu)\right] \leq \frac{u_f(n)-1}{2} \cdot \frac{\log n}{n} + \underbrace{\frac{ 2+ \log e}{n}}_{O(1/n)}. 
\end{align}
It is important to note that the bound in the RHS of (\ref{eq_main_proof_pre_34})  is valid independent of $k$. 

\subsubsection{Maximum Entropy analysis over the Envelope Family}
For what follows, let us consider the assumption that\footnote{This is equivalent to the condition $\Lambda_f  \subset \mathcal{H}(\mathbb{X})$ used in statement of Theorem \ref{main_th_part1} --- part iii).}
\begin{align}\label{eq_main_proof_pre_36}
	\sup_{\mu\in \Lambda_f }H(\mu) < \infty.
\end{align}
%
Then from (\ref{eq_main_proof_pre_34}),  we have that there is a coding scheme $ \left\{ \tilde{\mathcal{C}}^{*k}_n, n\geq 1 \right\}$  satisfying  that $\forall k\geq 1$: 
\begin{align}\label{eq_main_proof_pre_35}
	 \sup_{\mu \in \Lambda_f}  \frac{1}{n}   \mathbb{E}_{X^n \sim \mu^n}  \left\{ \mathcal{L}(\tilde{\mathcal{C}}^{*k}_n(O_k(X^n))) \right\}  &\leq \frac{u_f(n)-1}{2}  \frac{\log n}{n} + O(1/n) +  
	\sup_{\mu\in \Lambda_f }H(m_{\mu,k}),\nonumber\\
	    &= \frac{u_f(n)-1}{2}  \frac{\log n}{n} + O(1/n) +  \sup_{\mu\in \Lambda_f }H_{\sigma(\tilde{\pi}_k)}(\mu),\ \forall n\geq 1, 
\end{align}
where in the first inequality $m_{\mu,k} \in \mathcal{P}(\left\{1 \right\} \cup \Gamma_k^c)$ denotes the distribution of $Z=O_k(X)$ when $X\sim \mu\in \Lambda_f$, and in the second inequality, we use the tail partition $\tilde{\pi}_k = \left\{ \Gamma_k, \left\{ k+1\right\}, \left\{ k+2\right\}, \ldots \right\}$.
To continue with the argument, we use Lemma \ref{pro_maximun_entropy_tail_family} that shows that $\tilde{\mu}_f$ in (\ref{eq_main_proof_pre_11})
achieves the maximum entropy of the problem stated in the right term of (\ref{eq_main_proof_pre_35}) (eventually in $k$).  
Then assuming (\ref{eq_main_proof_pre_36}), i.e., $H(\tilde{\mu}_f)<\infty$, and a sufficiently large $k$, 
\begin{align}\label{eq_main_proof_pre_37}
	\frac{1}{n}   \sup_{\mu \in \Lambda_f} \mathbb{E}_{X^n \sim \mu^n}  \left\{ \mathcal{L}(\tilde{\mathcal{C}}^{*k}_n(O_k(X^n))) \right\}  &\leq \frac{u_f(n)-1}{2}  \frac{\log n}{n} + O(1/n)\nonumber\\
	 &+ \tilde{\mu}_f(\Gamma_k) \log \frac{1}{\tilde{\mu}_f(\Gamma_k)} + \sum_{i\geq k+1} \tilde{\mu}_f(i) \log \frac{1}{\tilde{\mu}_f(i)}, \ \forall n \geq 1. 
\end{align}

\subsubsection{Concatenating the results in (\ref{eq_main_proof_pre_21})}  
From the expressions in  (\ref{eq_main_proof_pre_26}),  (\ref{eq_main_proof_pre_37}) and (\ref{eq_main_proof_pre_21}), 
we have that for any distortion $d>0$ and threshold $k\geq 1$, there is a two-stage scheme $\left\{ \mathcal{T}^{*k}_n=(\xi^{*k}_n,(\tilde{\mathcal{C}}^{*k}_n, \tilde{\mathcal{D}}^{*k}_n)), n\geq 1\right\}$ where  $\xi^{*k}_n=(\phi^{*k}_n,\mathcal{C}^{*k}_n, \mathcal{D}^{*k}_n)$ is the $D$-semifaithful code of the first stage,  operating at distortion $d$ with respect to $\left\{\rho^k_n, n\geq 1 \right\}$,  and $(\tilde{\mathcal{C}}^{*k}_n, \tilde{\mathcal{D}}^{*k}_n)$ is the variable-length encoder-decoder pair of the second stage, such that for any $n\geq 1$: 
\begin{align}\label{eq_main_proof_pre_38}
 \sup_{\mu \in \Lambda_f} \left[  \frac{1}{n} \mathbb{E}_{X^n\sim \mu^n} \left\{ \mathcal{L}(\mathcal{T}^{*k}_n(X^n)) \right\}  - \mathcal{R}_n(d,\mu^n)  \right]  &\leq  \frac{k \log(n+1)}{n} + \frac{u_f(n)-1}{2}  \frac{\log n}{n} + O(1/n)  \nonumber\\
 						&+ \tilde{\mu}_f(\Gamma_k) \log \frac{1}{\tilde{\mu}_f(\Gamma_k)} + \sum_{i\geq k+1} \tilde{\mu}_f(i) \log \frac{1}{\tilde{\mu}_f(i)},
\end{align}
assuming 
that $H(\tilde{\mu}_f)<\infty$. Finally it is clear  in the above construction that we can take 
$(k_n)$ function of $n$ to achieve minimax universality using the fact that $(u_f(n) \cdot \log n/n)$ tends to zero with $n$ \cite{bontemps_2014,boucheron_2009}. 
In fact, if $(k_n)$ tends to $\infty$ with $n$ and 
$\lim_{n \longrightarrow \infty } k_n \log(n)/n=0$, from (\ref{eq_main_proof_pre_38}) this it is sufficient  to have that:  
\begin{align}\label{eq_main_proof_pre_39}
	\lim_{n \longrightarrow \infty }  \sup_{\mu \in \Lambda_f} \left[ \frac{1}{n} \mathbb{E}_{X^n\sim \mu^n} \left\{ \mathcal{L}(\mathcal{T}^{*k_n}_n(X^n)) \right\}  - \mathcal{R}_n(d,\mu^n)  \right] =0. 
\end{align}
Consequently, we achieve strong-minimax universality with the construction $\left\{ \mathcal{T}^{*k_n}_n, n\geq 1 \right\}$ in the sense stated in  (\ref{eq_sub_sec_UdSC_1}). This concludes the proof of  Part iii). 
\end{proof}

\subsection{Theorem  \ref{main_th_part1} --- Part ii): $f\in \ell_1(\mathbb{X})$ and $H(\tilde{\mu}_f)=\infty$}
\begin{proof}
	If we relax the finite entropy condition on the envelope distribution, i.e., $H(\tilde{\mu}_f)=\infty$,  the same 
	arguments, and in particular the two-stage construction presented in Section \ref{sec_main_proof_achievability} 
	can be used 
	to show that for any $\mu\in \Lambda_f$, such that $H(\mu)<\infty$, it follows that\footnote{For sake of space, the steps to derive (\ref{eq_main_proof_pre_40})  
	are not presented  as it follows directly from Section \ref{sec_main_proof_achievability}.} 
	\begin{align}\label{eq_main_proof_pre_40}
		\underbrace{ \left[ \frac{1}{n} \mathbb{E}_{X^n\sim \mu^n} \left\{ \mathcal{L}(\mathcal{T}^{*k_n}_n(X^n)) \right\}  - \mathcal{R}_n(d,\mu^n)  \right]}_{\text{point-wise analysis}}  &\leq  \frac{k_n \log(n+1)}{n} + \frac{u_f(n)-1}{2}  \frac{\log n}{n} + O(1/n)  \nonumber\\
 						&+ {\mu}(\Gamma_k) \log \frac{1}{{\mu}(\Gamma_k)} + \sum_{i\geq k+1}{\mu}(i) \log \frac{1}{{\mu}(i)}.
	\end{align}
	Then under the conditions that $(1/k_n)$ is $o(1)$ and $(k_n)$ is $o(\log(n)/n)$,  for 
	any $\mu \in \Lambda_f\cap H(\mathbb{X})$  it follows that
	\begin{align}\label{eq_main_proof_pre_41}
	 \lim_{n \longrightarrow \infty }   \left[  \frac{1}{n} \mathbb{E}_{X^n\sim \mu^n} \left\{ \mathcal{L}(\mathcal{T}^{*k_n}_n(X^n)) \right\}  - \mathcal{R}_n(d,\mu^n)  \right] =0, 
	\end{align}
	 which concludes the proof of Part ii). 
\end{proof}

\subsection{Theorem  \ref{main_th_rate_overhead_envelope} } 
\begin{proof}
Let us consider the assumption that
\begin{align}\label{eq_main_proof_pre_42}
		\lim \sup_{k \rightarrow  \infty} \frac{ \sum_{i\geq k} \tilde{\mu}_f(i) \log (1/ \tilde{\mu}_f(i)) }{  \tilde{\mu}_f(\mathcal{T}_k) \log 1/ \tilde{\mu}_f(\mathcal{T}_k)} < \infty.
\end{align}	
Notice that the expression in the numerator  is well defined when $H(\tilde{\mu}_f)<\infty$. 
Hence,  the result in (\ref{eq_main_proof_pre_38}) for the worse-case overhead can be adopted. 
Using a sequence $(k_n)_n$  such that $k_n \rightarrow \infty$
then the term  $H_{\sigma(\tilde{\pi}_{k_n})}(\tilde{\mu}_f)$ in (\ref{eq_main_proof_pre_38}) can be expressed (in the limit) by:
\begin{align}\label{eq_main_proof_pre_43}
	\lim \sup_{n \rightarrow  \infty} H_{\sigma(\tilde{\pi}_{k_n})} (\tilde{\mu}_f) =  \lim \sup_{n \rightarrow \infty}  \tilde{\mu}_f(\mathcal{T}_{k_n}) \log 1/ \tilde{\mu}_f(\mathcal{T}_{k_n})  \left[ 1 + \frac{ \sum_{i\geq k_n} \tilde{\mu}_f(i) \log (1/ \tilde{\mu}_f(i)) }{  \tilde{\mu}_f(\mathcal{T}_{k_n}) \log 1/ \tilde{\mu}_f(\mathcal{T}_{k_n})} \right],
\end{align}
where from (\ref{eq_main_proof_pre_42}), there are two constants $K_0>0$ and $N>0$, such  that for any 
$n\geq N$: 
\begin{align}\label{eq_main_proof_pre_44}
	H_{\sigma(\tilde{\pi}_{k_n})} (\tilde{\mu}_f) &= {\mu}(\Gamma_{k_n}) \log \frac{1}{{\mu}(\Gamma_{k_n})} + \sum_{i\geq k_n+1}{\mu}(i) \log \frac{1}{{\mu}(i)} \nonumber\\
	&\leq  \tilde{\mu}_f(\mathcal{T}_{k_n}) \log 1/ \tilde{\mu}_f(\mathcal{T}_{k_n}) \cdot K_0. 
\end{align}
In particular, choosing  $(k^f_{n})_n =  (u_f(n))_n$ by the definition in (\ref{eq_main_proof_pre_33b}) it follows that:  
$\tilde{\mu}_f(\mathcal{T}_{k^f_n+1})<1/n$ and $\tilde{\mu}_f(\mathcal{T}_{k^f_n}) \geq 1/n$.  Then 
for any $n\geq 1$:
\begin{align}\label{eq_main_proof_pre_45}
\tilde{\mu}_f(\mathcal{T}_{k^f_n}) \log 1/ \tilde{\mu}_f(\mathcal{T}_{k^f_n}) \leq \frac{1}{n} \log n.
\end{align}
Therefore considering the two-stage scheme $\left\{ \mathcal{T}^{*k^f_n}_n, n\geq 1 \right\}$  driven by $(k^f_{n})_{n\geq 1}$, 
from (\ref{eq_main_proof_pre_38}),  (\ref{eq_main_proof_pre_44}) and (\ref{eq_main_proof_pre_45}), we have that  
eventually in $n$
\begin{align}\label{eq_main_proof_pre_46}
 	\sup_{\mu \in \Lambda_f} \left[  \frac{1}{n} \mathbb{E}_{X^n\sim \mu^n} \left\{ \mathcal{L}(\mathcal{T}^{*k^f_n}_n(X^n)) \right\}  - \mathcal{R}_n(d,\mu^n)  \right]  	&\leq  \frac{u_f(n) \log(n+1)}{n} + \frac{u_f(n)-1}{2}  \frac{\log n}{n}\nonumber\\
		&+ O(1/n) +  K_0 \cdot \frac{\log n}{n},
\end{align}
which concludes the proof. 
\end{proof}

\section{Discussion and Concluding Remarks}
\label{final}
On the general analysis of universal $D$-semifaithful coding presented in Section \ref{sub_sec_UdSC} of this work, Theorem \ref{th_necessary_suffuficient_cond_universality} tells us that meeting minimax universality for a given non-zero distortion $d>0$ and a family of distributions $\Lambda$ implies the existence of a universal sequence of $D$-semifaithful quantizers for $\Lambda$. Consequently,  if the minimax redundancy criterion in (\ref{eq_sub_sec_UdSC_1}) is met,  for some $d>0$,  then there exists a sequence of partitions $\left\{ \pi_n, n\geq 1 \right\}$, such that $\pi_n\in \mathcal{Q}_n(d)$ (introduced in (\ref{eq_sec_pre_5b})), satisfying that 
\begin{equation}
\label{eq_final_1}
\lim_{n \rightarrow \infty} \frac{1}{n}  \sup_{\mu^n \in  \Lambda^n_f} \left[ H_{\sigma({\pi_n})}(\mu^n) - \min_{\pi \in   \mathcal{Q}_n(d)}   H_{\sigma(\pi)}(\mu^n)  \right]=0,
\end{equation}
where $H_{\sigma({\pi_n})}(\mu^n)$ is the entropy of $\mu^n$ restricted to the sub-sigma field induced by $\pi_n$ (see Eq.(\ref{eq_sec_pre_3})), and 
$\min_{\pi \in \mathcal{Q}_n(d)}   H_{\sigma(\pi)}(\mu^n)$  is  the  quantizer in $\mathcal{Q}_n(d)$ that minimizes the entropy 
given the distribution $\mu^n$ and $d$.  For obvious reasons, this representation dimension of the problem in (\ref{eq_final_1})  
 is not part of the lossless setting and requires a special treatment in this lossy case.  
In principle,  it is not obvious that the criterion in (\ref{eq_final_1}) can be achieved for any family of stationary memoryless distributions in $\infty$-alphabets. On this, a direct implication of Theorem \ref{main_th_part1} for envelope families (the achievability part in iii))  is that there is a universal quantization scheme in the sense presented in (\ref{eq_final_1}) for $\Lambda_f$ when $f\in \ell_1(\mathbb{X})$.  The proof of Theorem \ref{main_th_part1} in Section \ref{sec_main_proof_achievability} offers a concrete construction  for this universal quantization scheme based on the two-stage quantization approach illustrated in Figure \ref{fig1}.

On the analysis of universal $D$-semifaithful source coding on envelope families, Theorem \ref{main_th_part1} offers a necessary and sufficient condition to achieve  minimax universality (in the sense introduced in Section \ref{sub_sec_UdSC})  for  
$\Lambda_f$ in $\infty$-alphabets.  Interestingly, the condition matches the summability condition over $f$ known for the lossless (variable length) coding setting \cite{boucheron_2009}. 
 
Finally, it remains an open problem to evaluate if the rate of convergence for the worse-case overhead obtained  in Theorem \ref{main_th_rate_overhead_envelope} can be improved.  It is intriguing that this result does not show a faster rate of convergence to zero with $n$ (because of the non-zero distortion) with respect to its lossless counterpart that has the same rate. In fact, the  result is insensitive to the value of $d$, which is something that requires a more careful analysis.  In favor of the potential tightness of this part, we note that the non-zero distortion did not show an effect on the impossibility part (part i) of Theorem \ref{main_th_part1}) with respect to its counterpart in the lossless problem \cite{boucheron_2009}.  On the other hand, it is clear that the distortion reduces the information radius of the projected family, in the sense that 
$R^+(\Lambda_f^n, \sigma({\pi_n})) \leq R^+(\Lambda_f^n)$ (see the definition in  Eq.(\ref{eq_main_proof_pre_3b})).  
Then, the non-zero distortion does reduce this information radius complexity indicator. However, it is unclear that this gain in information radius translates into a gain in the overall minimax overhead expression in the lossy setting (with respect to its counterpart in the lossless setting) because  the information radius captures only one the two expressions of the redundancy in (\ref{eq_main_proof_pre_1}). The other non-negative term is captured by the role of the universal quantization discrepancy mentioned in (\ref{eq_final_1}).  

To conclude this discussion, we realize (from the expression in (\ref{eq_main_proof_pre_1}) and the analysis in Section \ref{sec_proof_main_pre_information_radius}) that  a concrete way to prove that the result in Theorem \ref{main_th_rate_overhead_envelope} is optimal is to show that any sequence of partitions  $\left\{ \pi_n, n\geq 1 \right\}$ such that $\pi_n\in \mathcal{Q}_n(d)$ satisfies that
\begin{equation}
\label{eq_final_2}
\lim \inf_{n \rightarrow \infty} \frac{R^+(\Lambda_f^n, \sigma(\pi_n))}{R^+(\Lambda_f^n)} >0.  
\end{equation}
At a first glance, this result looks not very intuitive, but we could conjecture that it is true. Indeed, a related non-zero gain (information radius) result has been obtained by the authors of this work  in 
\cite{silva_2017,silva_piantanida_reprint_2017}  but in a simpler context involving a tail-based scalar quantization and a distortion  that is not fixed and tends to $0$ with $n$.  We believe  that some of the tools used in this analysis can be adopted to derive (\ref{eq_final_2}), but the extension to analyze the object in (\ref{eq_final_2}) is not direct.  This is definitely a relevant direction for future work on universal source coding on $\infty$-alphabet.

\section{Acknowledgment}
The work of J.F. Silva was supported by Fondecyt 1210315 CONICYT-Chile and the Advanced Center for Electrical and Electronic Engineering, Basal Project FB0008. The work of Prof. Pablo Piantanida was supported by the European Commission’s Marie Sklodowska-Curie Actions (MSCA), through the Marie Sklodowska-Curie IF (H2020-MSCAIF-2017-EF-797805).

\appendices

\section{Proof of  Lemma \ref{prop_polinomial_envelop}}
\label{proof_prop_polinomial_envelop}
\begin{proof} 
	First, it is simple to verify that if $p>1$, then $(f_p(i)\log 1/ f_p(i))_{i \geq 1} \in \ell_1(\mathbb{X})$, 
	which implies that $\tilde{\mu}_{f_p}\in \mathcal{H}(\mathbb{X})$ (see Eq.(\ref{eq_main_proof_pre_11})).
	Let us introduce the tail series: 
	$$\mathcal{S}^k_p \equiv \sum_{i\geq k} \tilde{\mu}_{f_p}(i)=  \sum_{i\geq k}{f_p}(i),$$ 
	where the last equality is valid eventually (for $k$ sufficiently large).  Then it follows that: 
	\begin{align}\label{eq_prop_polinomial_envelop_1}
		\mathcal{S}^k_p= k^{-p} \sum_{i\geq k} \frac{k^p}{i^p}&=  k^{-p} \left( 1+ \frac{1}{((k+1)/k)^p} + \frac{1}{((k+2)/k)^p} +  \ldots  \frac{1}{((k+K)/k)^p}+  \ldots  \right)\nonumber\\
		 &= k^{-p} \left(1 + \sum_{i\geq 1} \frac{1}{(1+i/k)^p}  \right). 
	\end{align}
      The term of the series in the bracket  in the RHD of (\ref{eq_prop_polinomial_envelop_1}) is indexed by the fraction $i/k$,  
      where $k$ is fixed and $i$ goes over the integers.  Hence, this series decomposes in $k$-additive components as follows:
	\begin{align}\label{eq_prop_polinomial_envelop_2}
		\underbrace{\left(1 + \sum_{i\geq 1} \frac{1}{(i+1)^p}  \right)}_{\text{term with $0$ offset}} +  \underbrace{ \sum_{i\geq 1} \frac{1}{(i+1/k)^p}}_{\text{term with $1/k$ offset}}+ \ldots + \underbrace{\sum_{i\geq 1} \frac{1}{(i+(k-1)/k)^p}}_{\text{term with $(k-1)/k$ offset}}.
	\end{align}
	The $0$-offset term  in (\ref{eq_prop_polinomial_envelop_2}) equals  $\sum_{i\geq 1} \frac{1}{i^p}=S_p^1$. The $l/k$-offset term is upper bounded by $\sum_{i\geq 1} \frac{1}{i^p}=S_p^1$ and lower bounded by $ \sum_{i\geq 1} \frac{1}{(i+1)^p}=\sum_{i\geq 2} \frac{1}{i^p}=S_p^2$ for any $l\in \left\{1,..,k-1 \right\}$. Therefore from (\ref{eq_prop_polinomial_envelop_1}) and (\ref{eq_prop_polinomial_envelop_2}), we have that
	\begin{align}\label{eq_prop_polinomial_envelop_3}
		 \frac{1}{k^{p-1}} S^1_p  \geq S^k_p \geq  \frac{1}{k^p}  \left( S_p^1 + (k-1) S^2_p \right) \geq  \frac{1}{k^{p-1}} S^2_p, 
	\end{align}
which means that $S^k_p \sim \frac{1}{k^{p-1}}$.  When $p>1$, this term  tends to zero with $k$.

To continue with the proof, let us analyze the information series:  
	$$I^k_p \equiv \sum_{i\geq k} \tilde{\mu}_{f_p}(i) \log (1/\tilde{\mu}_{f_p}(i)) =  \sum_{i\geq k}  {f_p}(i) \log (1/{f_p}(i)),$$  
where the last equality is valid eventually (for $k$ sufficiently large). This last expression is equal to $p \sum_{i\geq k} \frac{1}{i^p} \log i$.  Therefore, we can  concentrate on the series:
	\begin{align}\label{eq_prop_polinomial_envelop_4}
		\tilde{I}^k_p  \equiv \sum_{i\geq k} \frac{1}{i^p} \log i &= \frac{\log k}{k^p} \left[ 1+ \sum_{i\geq 1} \frac{\log(k+i)/ \log(k)}{((k+i)/k)^p}  \right]	\nonumber\\
											&= \frac{\log k}{k^p} \left[ 1+ \sum_{i\geq 1} \frac{\log(k+i)/ \log(k)}{(1+i/k)^p}  \right].
	\end{align}
	Similarly to (\ref{eq_prop_polinomial_envelop_2}), the series in the RHD of (\ref{eq_prop_polinomial_envelop_4}) can be decomposed 
	in: 
	\begin{align}\label{eq_prop_polinomial_envelop_5}
		\underbrace{\left[ 1+ \sum_{i\geq 1} \frac{\log(k+ki)/ \log(k)}{(1+i)^p}  \right] }_{\text{$0$-term}} + \underbrace{\sum_{i\geq 1} \frac{\log(ik +1)/ \log(k)}{(i+1/k)^p}}_{\text{$1/k$-offset term}} + \ldots + \underbrace{\sum_{i\geq 1} \frac{\log(ik+k-1)/ \log(k)}{(i+(k-1)/k)^p}}_{\text{$(k-1)/k$-offset term}}.
	\end{align}
	For the $0$-offset term, we have that:
	\begin{align}\label{eq_prop_polinomial_envelop_6}
		\left[ 1+ \sum_{i\geq 1} \frac{\log(k+ki)/ \log(k)}{(1+i)^p}  \right] &\leq 1+  \sum_{i\geq 1}  \left( \frac{1}{1+i}\right)^p + \frac{1}{\log k} \sum_{i\geq 1} \frac{\log (i+1) }{(i+1)^p} \nonumber\\
		&= S^1_p + \frac{1}{\log k} I^2_p, 
	\end{align}
	while for the generic $l/k$-term in (\ref{eq_prop_polinomial_envelop_5}), we have that:
	\begin{align}\label{eq_prop_polinomial_envelop_7}
		\sum_{i\geq 1} \frac{\log(ik+l)/ \log(k)}{(i+l/k)^p}  &\leq  \sum_{i\geq 1} \frac{\log(ik+k)/ \log(k)}{i^p} \nonumber\\
		&= \sum_{i\geq 1} \frac{1}{i^p}  + \frac{1}{\log k} \underbrace{\sum_{i\geq 1}\frac{\log (i+1)}{i^p}}_{\bar{I}_p \equiv} 
			\nonumber\\ &=S_p^1 + \frac{1}{\log (k)} \bar{I}_p.
	\end{align}
	Returning to (\ref{eq_prop_polinomial_envelop_4}), it follows from (\ref{eq_prop_polinomial_envelop_5}) and the posterior bounds that
	\begin{align}\label{eq_prop_polinomial_envelop_8}
		{I}^k_p  \leq \frac{p \log k}{k^{p-1}}  \left[ S_p^1  + \frac{1}{\log k} \bar{I}_p  \right].
	\end{align}
	Then, 
	\begin{align}\label{eq_prop_polinomial_envelop_9}
		\lim \sup_{k \rightarrow  \infty} \frac{{I}^k_p}{S_p^k \log (1/S_p^k)}& \leq  \lim \sup_{k \rightarrow  \infty} \frac{\frac{p \log k}{k^{p-1}}  S_p^1  + \frac{p}{k^{p-1}} \bar{I}_p }{\frac{1}{k^{p-1}} S^2_p  \log \frac{k^{p-1}}{S_p^1}}\\
		& = \frac{p S_p^1}{(p-1)S^2_p} < \infty, 
	\end{align}
	which concludes the proof as $p>1$. 
\end{proof}

\section{Proof of Lemma \ref{prop_exponential_envelop}}
\label{proof_prop_exponential_envelop}
\begin{proof}
	If we consider the information function $(i_\alpha(i))=(- f_\alpha(i) \log f_\alpha(i))$, it is clearly summable then 
	$\tilde{\mu}_{f_\alpha} \in \mathcal{H}(\mathbb{X})$ (see Eq.(\ref{eq_main_proof_pre_11})). Let us analyze 
	the tail of $\tilde{\mu}_{f_\alpha}$, i.e., $S^k_\alpha \equiv \sum_{i\geq k} \tilde{\mu}_{f_\alpha}(i)$
	for any $k\geq 1$.
		We have that $S^k_\alpha=  e^{-\alpha k} \sum_{i\geq 1} K e^{-\alpha i}$ $= e^{-\alpha k} \cdot S^1_\alpha$.
	On the other hand, we need to analyze the tail fraction of the entropy of $\tilde{\mu}_{f_\alpha}$, i.e.,  
	$I^k_\alpha \equiv - \sum_{i \geq k}  \tilde{\mu}_{f_\alpha}(i) \log \tilde{\mu}_{f_\alpha}(i)$  
	$= - \sum_{i \geq k} f(i) \log f(i)$,  the last equality  holding eventually (for $k$ sufficiently large).
	It is simple to show that
	\begin{align}\label{eq_prop_exponential_envelop_1}
	  	I^k_\alpha= \log ({1}/{K}) S^k_\alpha + K\alpha \log e \cdot  \underbrace{\sum_{i\geq k} i e^{-\alpha i} }_{\bar{I}^k_\alpha \equiv}
	\end{align}
	where  $\bar{I}^k_\alpha= k e^{-\alpha k} S^0_\alpha  \left(1/K + 1/k \cdot e^{-\alpha} \right)$.  Finally,  we have from (\ref{eq_prop_exponential_envelop_1}) that  $I^k_\alpha= (\log ({1}/{K}) S^1_\alpha + S^1_\alpha) \cdot e^{-\alpha k}  + (S^0_\alpha/K) \cdot k e^{-\alpha k}$.  With this,  it is simple to verify that:
	\begin{align}\label{eq_prop_exponential_envelop_2}
		\lim \sup_{k \rightarrow  \infty} \frac{I^k_\alpha}{S_\alpha^k \log (1/S_\alpha^k)}& = \frac{S^0_\alpha}{K S^1_\alpha}  \cdot \lim \sup_{k \rightarrow  \infty} \frac{ k}{ k \alpha \log e+ \log (1/S^1_\alpha)} \nonumber\\
		&=\frac{1}{K e^{-\alpha} \alpha \log e}   <  \infty,  
	\end{align}
	which proves the result. 
\end{proof}

\section{Proof of Lemma \ref{lemma_finite_alphabet_unievrsal_construction}}
\label{proof_lemma_finite_alphabet_unievrsal_construction}
\begin{proof}
	Without loss of generality, let us consider the finite alphabet $\mathcal{A}= \left\{1,..,k\right\}$, a distortion $d>0$, 
	and the collection $\Lambda=\mathcal{P}(\mathcal{A})$.  Using the non-asymptotic performance bound in (\ref{eq_sec_pre_5b}), 
	we are interested in the following object:
	\begin{equation} \label{eq_proof_lemma_finite_alphabet_unievrsal_construction_1}
		\min_{(\phi_n,\mathcal{C}_n,\mathcal{D}_n)}  \sup_{\mu \in \Lambda } \left[ \frac{1}{n}  \mathbb{E}_{X^n\sim \mu^n}  \left\{  \mathcal{L}(\mathcal{C}_n(\phi_n(X^n))) \right\} -   \mathcal{R}_n(d,\mu^n) \right], 
	\end{equation}
	where the minimum is carried over the collection of $D$-semifaithful codes on $\mathcal{A}$ operating at distortion $d$.

	Let us fix an arbitrary $\epsilon>0$. For any $x^n\in \mathcal{A}^n$, let $p_{x^n}$ denote the type of $x^n$ (the empirical distribution 
	in $\mathcal{P}(\mathcal{A})$ induced by $x^n$), and $\tilde{P}_n \equiv \left\{p_{x^n}, x^n \in  \mathcal{A}^n \right\}$ 
	the collection of types obtained with sequences of length $n$. For any $p\in \tilde{P}_n$,  the type class of $p$
	is given by $T_p \equiv \left\{x^n\in \mathcal{A}^n: p_{x^n}=p  \right\}$, where it is clear that $\left\{T_p, p\in \tilde{P}_n \right\}$ offers a finite partition of $\mathcal{A}^n$.
	It is well known that $ \left| \tilde{P}_n \right| \leq ({n+1)}^{k}$ \cite{cover_2006}.  For any member in the type class $p\in \tilde{P}_n$, let us 
	choose a $D$-semifaithful code $\xi^{*k}_{n,p}=(\phi^{*k}_{n,p},\mathcal{C}^{*k}_{n,p}, \mathcal{D}^{*k}_{n,p})$ indexed by $p$ satisfying the condition:\footnote{This selection can be accomplished from (\ref{eq_sec_pre_5b}).} 
	\begin{equation} \label{eq_proof_lemma_finite_alphabet_unievrsal_construction_2}
		\frac{1}{n}\mathbb{E}_{Y^n\sim \bar{\mu}_p}  \left\{  \mathcal{L}(\mathcal{C}^{*k}_{n,p}(\phi^{*k}_{n,p}(Y^n))) \right\}  \leq  \mathcal{R}_n(d, \bar{\mu}_p) + \epsilon,
	\end{equation}
	where $\bar{\mu}_p\in \mathcal{P}(\mathcal{A}^n)$ in (\ref{eq_proof_lemma_finite_alphabet_unievrsal_construction_2}) 
	is a short-hand for the uniform distribution over $T_p\subset \mathcal{A}^n$.
	
	With this,  we consider a simple two-stage universal strategy, inspired by the two-stage scheme used  in lossless  universal  source coding \cite{csiszar_2004}. 
	For encoding $x^n$ there is fixed-rate function $f_n:\tilde{P}_n \rightarrow \left\{0,1 \right\}^{k \lceil \log (n+1) \rceil }$ for indexing (encoding) the type of $x^n$, 
	and  conditioning on this information, the second-stage encodes $x^n$ lossily with $\xi^{*k}_{n,p_{x^n}}$. Then the variable length representation of $x^n$ operating 
	at distortion $d$ is given by $(f_n(p_{x^n}), \mathcal{C}^{*k}_{n, p_{x^n}}(\phi^{*k}_{n,p_{x^n}}(x^n)))\in \left\{0,1 \right\}^*$. From this construction, 
	it is simple to check that this scheme is a $D$--semifaithful code of $\mathcal{A}^n$ with respect to $\rho_n$.  
	
	Let us analyze its worse-case 
	overhead in $\Lambda$.  Let us consider $\mu\in \Lambda$, then if we denote by $\mathcal{T}^k_{n}=(f_n, (\xi^{*k}_{n, p}; p\in \tilde{P}_n))$ 
	the two-stage scheme and (with small abuse of notation) we use $\mathcal{T}^k_{n}$ as a short-hand for the encoding mapping (from source symbols
	to binary sequences) then: 
	\begin{equation}
	\label{eq_proof_lemma_finite_alphabet_unievrsal_construction_3}
	\mathcal{L}(\mathcal{T}^k_{n}(x^n))= \underbrace{k\log (n+1)}_{\text{first-stage}} + \underbrace{\mathcal{L}(\mathcal{C}^{*k}_{n, p_{x^n}}(\phi^{*k}_{n, p_{x^n}}(x^n))))}_{\text{second-stage}},\ \forall x^n \in \mathcal{A}^n
	\end{equation}
	and 
	\begin{align} 
		\frac{1}{n} \mathbb{E}_{X^n\sim \mu^n}  \left\{ \mathcal{L}(\mathcal{T}^k_{n}(X^n))  \right\}  &-  \mathcal{R}_n(d,\mu^n) 
		=  \frac{1}{n}  \mathbb{E}_ {Y\equiv T_{X^n}} \left\{  \mathbb{E}_{X^n| Y}  \left\{ \mathcal{L}(\mathcal{T}^k_{n}(X^n)) | Y \right\}  \right\} 
		-  \mathcal{R}_n(d,\mu^n)\nonumber\\
		\label{eq_proof_lemma_finite_alphabet_unievrsal_construction_3a}
		&=\frac{k\log (n+1)}{n} + \sum_{p\in \tilde{P}_n} \mu^n(T_p) \mathbb{E}_{X^n \sim \bar{\mu}_p} \left\{  \mathcal{L}(\mathcal{C}^{*k}_{n,p}(\phi^{*k}_{n,p}(X^n))) \right\}  -  \mathcal{R}_n(d,\mu^n)\\
		\label{eq_proof_lemma_finite_alphabet_unievrsal_construction_3b}
		&=\frac{k\log (n+1)}{n} +  \sum_{p\in \tilde{P}_n} \mu^n(T_p)   \left[ \frac{1}{n}  \mathbb{E}_{X^n \sim \bar{\mu}_p} \left\{ \mathcal{L}(\mathcal{C}^{*k}_{n,p}(\phi^{*k}_{n,p}(X^n))) \right\} -  \mathcal{R}_n(d, \bar{\mu}_p)  \right] \nonumber\\ 
		&+   \underbrace{ \sum_{p\in \tilde{P}_n} \mu^n(T_p)  \mathcal{R}_n(d, \bar{\mu}_p)  -  \mathcal{R}_n(d,\mu^n) }_{\leq 0}\\
		\label{eq_proof_lemma_finite_alphabet_unievrsal_construction_3c}
		&\leq  \frac{k\log (n+1)}{n} + \epsilon.
	\end{align}
	The expression in (\ref{eq_proof_lemma_finite_alphabet_unievrsal_construction_3a}) follows from
	(\ref{eq_proof_lemma_finite_alphabet_unievrsal_construction_3}) and the observation that conditioning to the event $Y=p$,  for some valid $p\in \tilde{P}_n$, 
	$X^n\sim \bar{\mu}_p$ independent of $\mu^n$ \cite{cover_2006}. To obtain (\ref{eq_proof_lemma_finite_alphabet_unievrsal_construction_3b}),  we include 
	the term $ \sum_{p\in \tilde{P}_n} \mu^n(T_p)  \mathcal{R}_n(d, \bar{\mu}_p) $ in (\ref{eq_proof_lemma_finite_alphabet_unievrsal_construction_3a}) 
	to then use the  inequality in (\ref{eq_proof_lemma_finite_alphabet_unievrsal_construction_2}).
	Finally to obtain (\ref{eq_proof_lemma_finite_alphabet_unievrsal_construction_3c}), we use the fact that $\mu^n(B)= \sum_{p\in \tilde{P}_n} \mu^n(T_p) \bar{\mu}_p (B)$
	\cite{cover_2006} and that $ \mathcal{R}_n(d,\mu)$ is a concave function of the second argument from its construction in (\ref{eq_sec_pre_5b}).
	Finally, the inequality in (\ref{eq_proof_lemma_finite_alphabet_unievrsal_construction_3c}) is valid distribution free, which concludes the proof. 
\end{proof}

\section{Proof of Lemma \ref{pro_maximun_entropy_tail_family}}
\label{proof_pro_maximun_entropy_tail_family}
\begin{proof} 
Let us assume that $H(\tilde{\mu}_f)<\infty$,  where $\tilde{\mu}_f\in \Lambda_f$ is the tail distribution introduced in (\ref{eq_main_proof_pre_11}).
Let us consider an arbitrary $\mu\in \Lambda_f$.  Then we have that (assuming the regime where $k> \tau_f$, see (\ref{eq_main_proof_pre_11})): 
\begin{align}
	\label{eq_pr_pro_maximun_entropy_tail_family_1}
 	&H_{\sigma(\tilde{\pi}_k)}(\tilde{\mu}_f) - H_{\sigma(\tilde{\pi}_k)}(\mu) = \nonumber\\ 
	&\tilde{\mu}_f(\Gamma_k) \log \frac{1}{\tilde{\mu}_f(\Gamma_k) } + \sum_{x\geq k+1} \mu(x) \log \frac{\mu(x)}{f(x)} +  \sum_{x\geq k+1} (f(x)- \mu(x)) \log \frac{1}{f(x)} - {\mu}(\Gamma_k) \log \frac{1}{{\mu}(\Gamma_k)} \nonumber\\
					&\geq \tilde{\mu}_f(\Gamma_k) \log \frac{1}{\tilde{\mu}_f(\Gamma_k) } +  \mu(\Gamma_k) \log \frac{\tilde{\mu}_f(\Gamma_k)}{\mu(\Gamma_k)} + {\mu}(\Gamma_k) \log {{\mu}(\Gamma_k)}\nonumber\\ 
					&+  \sum_{x\geq k+1} (f(x)- \mu(x)) \log \frac{1}{f(x)}\\
					\label{eq_pr_pro_maximun_entropy_tail_family_1b}
					&=(\tilde{\mu}_f(\Gamma_k)-  {\mu}(\Gamma_k)) \cdot \log \frac{1}{\tilde{\mu}_f(\Gamma_k) } +  \sum_{x\geq k+1} (f(x)- \mu(x)) \log \frac{1}{f(x)},\nonumber\\ 
					&= \left(  \sum_{x\geq k+1}{\mu}(x)- \sum_{x\geq k+1}{\tilde{\mu}_f}(x) \right) \cdot \log \frac{1}{\tilde{\mu}_f(\Gamma_k) } +  \sum_{x\geq k+1} (f(x)- \mu(x)) \log \frac{1}{f(x)}\nonumber\\
					&= \sum_{x\geq k+1} (f(x)- \mu(x)) \cdot \log \frac{1-\sum_{y\geq k+1}f(y) }{f(x)}.
\end{align}
To obtain  (\ref{eq_pr_pro_maximun_entropy_tail_family_1}) we use that $\sum_{x\geq k+1} \mu(x) \log \frac{\mu(x)}{f(x)}\geq - \mu(\Gamma_k) \log \frac{\mu(\Gamma_k)}{\tilde{\mu}_f(\Gamma_k)}$ from the observation that $D_{\sigma(\tilde{\pi}_k)}(\mu\|\tilde{\mu}_f) \geq 0$.
At this point, we use the fact that $f\in \ell_1(\mathbb{X})$, which means that $\lim_{k \rightarrow \infty} \sum_{x\geq k+1}f(x)=0$. 
Therefore eventually (i.e., for a sufficiently large $k$) we have that  $1- \sum_{x\geq k+1}f(x) > \sum_{x\geq k+1}f(x) $.  Assuming this 
large $k$ regime,  it follows from (\ref{eq_pr_pro_maximun_entropy_tail_family_1b}) that
\begin{align}
	\label{eq_pr_pro_maximun_entropy_tail_family_2}
	H_{\sigma(\tilde{\pi}_k)}(\tilde{\mu}_f) - H_{\sigma(\tilde{\pi}_k)}(\mu) & \geq  \sum_{x\geq k+1} (f(x)- \mu(x)) \cdot \log \frac{\sum_{y\geq k+1}f(y) }{f(x)} \geq 0. \end{align}
The last inequality in (\ref{eq_pr_pro_maximun_entropy_tail_family_2}) comes from the assumption that $\mu\in \Lambda_f$, which means that $\mu(x) \leq f(x)$ for all $x\in \mathbb{X}$.  

On the second part of the result, we assume that $H(\tilde{\mu}_f)=\infty$.  Here, it is clear that $H_{\sigma(\tilde{\pi}_k)}(\tilde{\mu}_f)=\infty$ for any $k\geq 1$, which is sufficient to obtain the unbounded result. 
 \end{proof}

\section{Proof of Lemma \ref{prop_infinite_information_radius}}
\label{proof_prop_infinite_information_radius}
\begin{proof}
	First, it is important to note that by the construction of 
	$\hat{\Lambda}^n$ in (\ref{eq_main_proof_pre_8})
	and the partition $\eta_n$ in (\ref{eq_main_proof_pre_9}), $\hat{\Lambda}^n$ degenerates in the probability 
	space $(\mathbb{X}^n, \sigma(\eta_n))$,  in the sense that for any $k\geq 1$
	\begin{equation} \label{eq_pr_prop_infinite_information_radius_1}
		H_{\sigma(\eta_n)}(\tilde{\mu}^n_{j_k})=0.
	\end{equation}
	Let us consider a distribution over the indices of the family $\hat{\Lambda}^n$ (i.e., over the integer set $\mathbb{N}$) $\rho\in \mathcal{P}(\mathbb{N})$, 
	and with this we can construct a joint distribution $\rho \times \hat{\Lambda}^n$ in the product space  $(\mathbb{N}, 2^{\mathbb{N}}) \times (\mathbb{X}^n, \sigma(\eta_n))$ in the standard way, i.e., $\rho \times \hat{\Lambda}^n(A\times B)=\sum_{a\in A} \rho(a)\cdot  \tilde{\mu}^n_{j_a}(B)$ for any $A\subset \mathbb{N}$ and $B\in \sigma(\eta_n)$. Associated with this joint distribution, we can derive an expression for the mutual information of $\rho \times \hat{\Lambda}^n$ \cite{cover_2006,csiszar_2004}: 
	\begin{align} \label{eq_pr_prop_infinite_information_radius_2}
		\mathcal{I}(\rho; \hat{\Lambda}^n) & \equiv  \sum_{a\in \mathbb{N}}\rho(a) \cdot D_{\sigma(\eta_n)}(\tilde{\mu}^n_{j_a}\|\bar{\mu}) \\
		&= H_{\sigma(\eta_n)}(\bar{\mu}) - \sum_{a\in \mathbb{N}} \rho(a) \cdot H_{\sigma(\eta_n)}(\tilde{\mu}^n_{j_a}), 
	\end{align}
	where $\bar{\mu}(B) \equiv  \sum_{a\in \mathbb{N}}\rho(a) \tilde{\mu}^n_{j_a}(B)$ for any $B\in \sigma(\eta_n)$. Using (\ref{eq_pr_prop_infinite_information_radius_1}), it is simple to show that $\mathcal{I}(\rho; \hat{\Lambda}^n)=H_{\sigma(\eta_n)}(\bar{\mu})= H(\rho)=-\sum_{a\in \mathbb{N}} \rho(a) \log \rho(a)$. 
	Finally it is well known,  from the construction of the information radius of $\hat{\Lambda}^n$ \cite{csiszar_2004}, that 
	$R^+(\hat{\Lambda}^n, \sigma(\eta_n)) \geq \mathcal{I}(\rho; \hat{\Lambda}^n)= H(\rho)$ for any $\rho\in \mathcal{P}(\mathbb{N})$. 
	This last inequality proves the result as $\sup_{\rho\in \mathcal{P}(\mathbb{N})}H(\rho)=\infty$.
\end{proof}
 
\section{Proposition \ref{prop_distortion_ineq}}
\label{proof_prop_distortion_ineq}
\begin{proposition}\label{prop_distortion_ineq}
For all $x^n\in \mathbb{X}^n$, it follows that $\rho_n(x^n, \hat{x}^n) \leq \tilde{\rho}_n(y^n, \hat{y}^n)$.
\end{proposition}
\begin{proof}
	\begin{align}\label{eq_prop_distortion_ineq_1}
		\rho_n(x^n, \hat{x}^n)& = \frac{1}{n} \sum_{i=1}^n \rho( x_i, \hat{x}_i) \nonumber\\
			&= \frac{1}{n} \sum_{i=1}^n  \left[ \rho( x_i, \hat{x}_i) {\bf 1}_{\Gamma_k}(x_i) + \rho( x_i, \hat{x}_i) {\bf 1}_{\Gamma_k^c}(x_i) \right]\nonumber\\
			&= \frac{1}{n} \sum_{i=1}^n \left[{\rho}(y_i, \hat{y}_i) {\bf 1}_{\Gamma_k}(x_i) + \underbrace{\rho(x_i, z_i)}_{=0 \text{ as } z_i=x_i} {\bf 1}_{\Gamma_k^c}(x_i) \right]\nonumber\\
			&\leq \frac{1}{n} \sum_{i=1}^n \tilde{\rho}( y_i, \hat{y}_i) {\bf 1}_{\Gamma_k}(x_i) \\
			&  \leq  \frac{1}{n} \sum_{i=1}^n  \tilde{\rho}(y_i, \hat{y}_i) \\
			& = \tilde{\rho}_n(y^n, \hat{y}^n).
	\end{align}
	The first inequality in (\ref{eq_prop_distortion_ineq_1}) follows from the construction of $\tilde{\rho}$ assuming that coincides with $\rho$
	in $\Gamma_k \times \Gamma_k$ and the mild assumption that 
	$\tilde{\rho}(i, k+1) \leq  {\rho}(i, k+1)$ for all $ i \in \Gamma_k$.
\end{proof}

\bibliographystyle{IEEEtran}				
\bibliography{main_jorge_silva}				

\end{document}